%% file: main_elsarticle.tex
\documentclass[preprint,12pt]{elsarticle}
\input{env_elsarticle}

\journal{arXiv}

\begin{document}

\begin{frontmatter}

%% Article title
\title{\textbf{A Multi-Component, Multi-Physics Computational Model for Solving Coupled Cardiac Electromechanics and Vascular Haemodynamics}}

%% Author name
\author[1]{Sharp C. Y. Lo}
\author[2]{Alberto Zingaro}
\author[1,3]{Jon W. S. McCullough}
\author[1]{Xiao Xue}
\author[2,4]{Pablo Gonzalez-Martin}
\author[5,6]{Balint Joo}
\author[2,7]{Mariano V\'azquez}
\author[1,8,9]{Peter V. Coveney}
\ead{p.v.coveney@ucl.ac.uk}

\affiliation[1]{
            organization={Centre for Computational Science, University College London},
            city={London},
            country={United Kingdom}}
\affiliation[2]{
            organization={ELEM Biotech SL},
            %addressline={Pier07 - Via Laietana, 26},
            %postcode={08003},
            city={Barcelona},
            country={Spain}}
\affiliation[3]{
            organization={School of Mechanical and Aerospace Engineering, Queen's University Belfast},
            city={Belfast},
            country={United Kingdom}}
\affiliation[4]{
            organization={PhySense, Department of Information and Communication Technologies, Universitat Pompeu Fabra},
            %addressline={Tànger, 122-140},
            %postcode={08018},
            city={Barcelona},
            country={Spain}}
\affiliation[5]{
            organization={NVIDIA Corporation},
            %addressline={2788 San Tomas Expressway},
            city={Santa Clara},
            %postcode={CA 95051},
            country={United States of America}}
\affiliation[6]{
            organization={National Center for Computational Science, Oak Ridge National Laboratory},
            %addressline={1 Bethel Valley Rd},
            city={Oak Ridge},
            %postcode={TN 37831},
            country={United States of America}}
\affiliation[7]{
            organization={Barcelona Supercomputing Center},
            %addressline={Plaça Eusebi G\"uell, 1-3},
            %postcode={08034},
            city={Barcelona},
            country={Spain}}
\affiliation[8]{
            organization={Advanced Research Computing Centre, University College London},
            city={London},
            country={United Kingdom}}
\affiliation[9]{
            organization={Informatics Institute, Faculty of Science, University of Amsterdam},
            city={Amsterdam},
            country={Netherlands}}

%% Abstract
\begin{abstract}
    The circulatory system, comprising the heart and blood vessels, is vital for nutrient transport, waste removal, and homeostasis. Traditional computational models often \rev{treat} cardiac electromechanics and blood flow dynamics \rev{separately, overlooking the integrated nature of the system}. This paper presents an innovative approach that couples a 3D electromechanical model of the heart with a 3D fluid mechanics model of vascular blood flow. \rev{Using a file-based partitioned coupling scheme,} these models run independently while sharing essential data through intermediate files. We validate this approach using solvers \rev{developed} by separate research groups, \rev{each targeting disparate} dynamical scales \rev{employing} distinct discretisation schemes, \rev{and implemented in different programming languages}. Numerical simulations using idealised and realistic anatomies show that the coupling scheme is reliable and requires minimal additional computation time relative to advancing individual time steps in the heart and blood flow models. Notably, the coupled model predicts muscle displacement \rev{and aortic wall shear stress} differently than the standalone \rev{models}, highlighting the \rev{importance of coupling between cardiac and vascular dynamics in cardiovascular simulations. Moreover, we demonstrate the model's potential for medical applications by simulating the effects of myocardial scarring on downstream vascular flow.} This study presents a paradigm case of how to build virtual human models and digital twins by productive collaboration between teams with complementary expertise.
\end{abstract}

\begin{graphicalabstract}
\includegraphics[trim = {0cm 0cm 0cm 0cm}, clip, width=\textwidth]{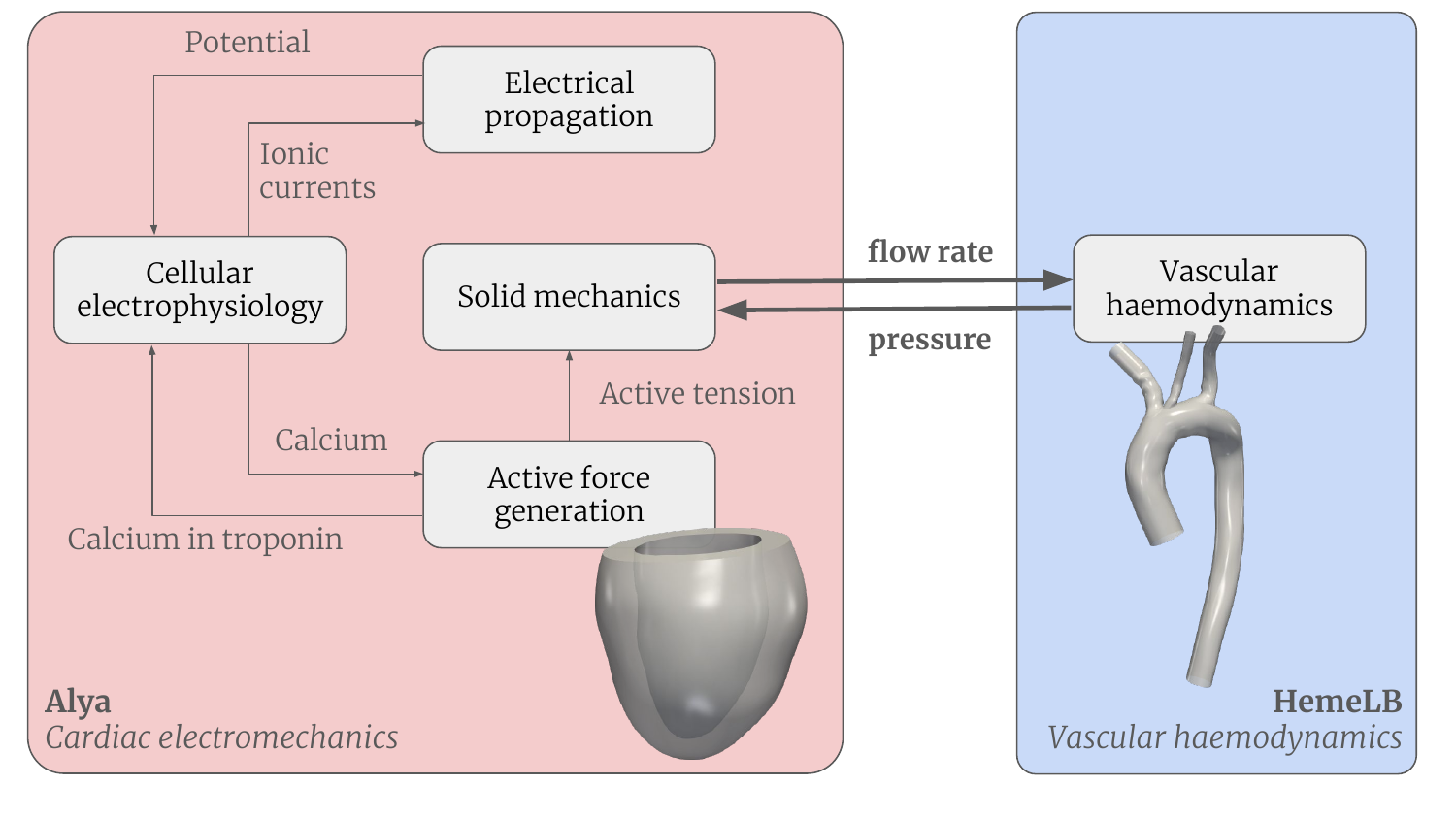}
\end{graphicalabstract}

\begin{highlights}
\item Innovative approach couples 3D cardiac and vascular simulations
\item Multi-component strategy leverages existing specialised solvers
\item Coupled model highlights influence of detailed vascular flow on cardiac function
\item File-based partitioned coupling scheme requires minimal extra computation time
\item Framework facilitates virtual human models and digital twins development
\end{highlights}

%% Keywords

\end{frontmatter}

\input{1_Introduction}
\input{2_Methods}
\input{3_Results-Discussion}
\input{4_Limitations}
\input{5_Conclusion}

\appendix
\input{6_Appendix}

\section*{CRediT Authorship Contribution Statement}
\textbf{Sharp C. Y. Lo:} Conceptualisation, Methodology, Software, Validation, Formal analysis, Investigation, Data curation, Writing --- original draft, Visualisation. \textbf{Alberto Zingaro:} Conceptualisation, Methodology, Software, Validation, Formal analysis, Investigation, Writing --- original draft, review \& editing, Visualisation, Supervision. \textbf{Jon W. S. McCullough:} Conceptualisation, Methodology, Software, Writing --- review \& editing, Funding acquisition. \textbf{Xiao Xue:} Software, Funding acquisition. \textbf{Pablo Gonzalez-Martin:} Methodology, Software. \textbf{Balint Joo:} Software, Writing --- review \& editing. \textbf{Mariano V\'azquez:} Conceptualisation, Writing --- review \& editing, Supervision, Project administration, Funding acquisition. \textbf{Peter V. Coveney:} Conceptualisation, Writing --- review \& editing, Supervision, Project administration, Funding acquisition.

\section*{Declaration of Competing Interest}
The authors declare that they have no known competing financial interests or personal relationships that could have appeared
to influence the work reported in this paper.

\section*{Declaration of Generative AI and AI-assisted Technologies in Scientific Writing}
During the preparation of this work the authors used ChatGPT to improve the readability and language of the manuscript. After using this tool, the authors reviewed and edited the content as needed and take full responsibility for the content of the published article.

\section*{Data Availability}
Supplementary materials including the vascular anatomies and the complete simulation and analysis data are available from the Figshare repository: \url{https://figshare.com/s/6a820c417ff707510000}.

\section*{Acknowledgements}
We acknowledge funding support from the European Commission CompBioMed Centre of Excellence (Grant No. 675451 and 823712) and the UK Engineering and Physical Sciences Research Council (EPSRC) under the projects ``UK Consortium on Mesoscale Engineering Sciences (UKCOMES)'' (Grant No. EP/R029598/1), ``Software Environment for Actionable \& VVUQ-evaluated Exascale Applications (SEAVEA)'' (Grant No. EP/W007711/1), and ``CompBioMedX: Computational Biomedicine at the Exascale'' (Grant No. EP/X019446/1). S. C. Y. L. is grateful for the research studentship funded by University College London and CBK Sci Con Ltd. A. Z. and M. V. acknowledge support from the European Union through the EIC Project No. 190134524, ``ELEM Virtual Heart Populations for Supercomputers'' (ELVIS), as well as the Horizon Europe research and innovation programme under grant agreement No. 101136728 (VITAL). Views and opinions expressed are, however, those of the authors only and do not necessarily reflect those of the European Union or EISMEA. Neither the European Union nor the granting authority can be held responsible for them. This work used the ARCHER2 UK National Supercomputing Service (\url{https://www.archer2.ac.uk}), made accessible via the EPSRC SEAVEA project, and resources of the Oak Ridge Leadership Computing Facility, which is a DOE Office of Science User Facility supported under Contract DE-AC05-00OR22725.

\bibliographystyle{elsarticle-num-names}
\bibliography{ref_Mendeley, ref_local}

\end{document}

%% file: env_elsarticle.tex
\usepackage[a4paper, left=2cm, right=2cm, top=2cm, bottom=2cm]{geometry}
\usepackage[english]{babel}
\usepackage{csquotes} % to ensure quoted texts are typeset when using babel with biblatex
\usepackage{subcaption}
\usepackage{physics}
\usepackage{amsmath, amssymb}
\usepackage{algorithm, algpseudocode}
\usepackage{enumitem}

\usepackage{hyperref}
\hypersetup{
    colorlinks=true,
    linkcolor=blue,
    urlcolor=blue,
    citecolor=blue,
}

\usepackage{xcolor}
\usepackage{listings}
\usepackage[per-mode=symbol,
            inter-unit-product=\cdot,
            separate-uncertainty=true,
            print-unity-mantissa=false,
            table-alignment-mode=none,
            table-alignment=right,
            exponent-product=\cdot]{siunitx}
\usepackage{booktabs}
\DeclareSIUnit\faraday{F}
\usepackage{mathtools,amsfonts,cases}
\usepackage{bm}
\usepackage{stackengine}[2021-07-15]
\usepackage[normalem]{ulem} % force linebreaks in biblatex

\newcommand{\rev}[1]{{\color{black}#1}}

\newcommand{\R}[1]{\mathbb R^{#1}}

\newcommand{\deformation}{F}

\newcommand{\displacement}{\bm d}
\newcommand{\gammaendo}{\Gamma^{\mathcal A}_{0_ \mathrm{endo}}}
\newcommand{\gammaplane}{\Gamma^{\mathcal A}_{0_ \mathrm{plane}}}
\newcommand{\gammapool}{\Gamma^{\mathcal A}_{0_ \mathrm{pool}}}
\newcommand{\gammaepi}{\Gamma^{\mathcal A}_{0_ \mathrm{epi}}}
\newcommand{\gammabase}{\Gamma^{\mathcal A}_{0_ \mathrm{base}}}
\newcommand{\domain}{\Omega^{\mathcal A}_0}
\newcommand{\domaint}{\Omega^{\mathcal A}_t}
\newcommand{\deformationoperator}{\bm \Phi}
\newcommand{\X}{\bm X}
\newcommand{\x}{\bm x}
\newcommand{\density}{\rho_0}
\newcommand{\FS}{\rev{FS(\displacement, \tact(\qact))}}

\newcommand{\normal}{\bm n_0}
\newcommand{\pendo}{p_\mathrm{endo}}
\newcommand{\JFmT}{J \deformation^{-\top}}
\newcommand{\kepi}{k_\mathrm{epi}}
\newcommand{\dert}[2]{\frac{\mathrm d^{#1}{#2}}{\mathrm d t^{#1}}}
\newcommand{\pmvo}{p_\mathrm{MVo}}

\newcommand{\pavo}{p_\mathrm{AVo}}
\newcommand{\pavc}{p_\mathrm{AVc}}

\newcommand{\D}{\bm D}
\newcommand{\condtens}{\bm \sigma}
\newcommand{\Iion}{I_\mathrm{ion}}
\newcommand{\Iapp}{I_\mathrm{app}}
\newcommand{\Cm}{C_\mathrm{m}}
\newcommand{\Sv}{S_\mathrm{v}}

\newcommand{\gating}{\bm w}
\newcommand{\concentration}{\bm c}
\newcommand{\condf}{\sigma_f}
\newcommand{\condn}{\sigma_n}
\newcommand{\conds}{\sigma_s}
\newcommand{\fzero}{\tilde{\bm f}_0}
\newcommand{\szero}{\tilde{\bm s}_0}
\newcommand{\nzero}{\tilde{\bm n}_0}

\newcommand{\tact}{T_\mathrm{act}}
\newcommand{\qact}{\bm q}
\newcommand{\Sact}{S_\mathrm{act}}
\newcommand{\Spass}{S_\mathrm{pass}}
\newcommand{\lambdaf}{\lambda_\mathrm{f}}

\newcommand{\domainheme}{\Omega^{\mathcal H}}
\newcommand{\gammainheme}{\Gamma_\mathrm{in}^\mathcal{H}}
\newcommand{\gammawallheme}{\Gamma_\mathrm{wall}^\mathcal{H}}
\newcommand{\gammaoutheme}{\Gamma_\mathrm{out}^\mathcal{H}}
\newcommand{\Qin}{Q_\mathrm{in}}

%% file: 1_Introduction.tex
\section{Introduction}
The circulatory system, comprising the heart and blood vessels, is a complex network that operates through a series of tightly coupled biophysical processes occurring at different levels. Key among these processes are electrophysiology, solid mechanics, and fluid dynamics. As an electrical impulse spreads through cardiomyocytes, calcium ions are released from intracellular stores, bind to troponin C, and trigger heart muscle contraction \cite{Trayanova2024ComputationalTranslation, 2024FamilialCardiomyopathies}. The increased pressure within the ventricles causes the atrioventricular valves to close and the semilunar valves to open, resulting in the ejection of blood from the ventricles into the arteries \cite{Pocock2018HumanPhysiology}. The circulation of blood in the vascular network, facilitated by vessel constriction and dilation, ensures nutrient transport, waste removal, and homeostasis \cite{Pocock2018HumanPhysiology}. Understanding these individual processes, as well as their interactions, is crucial for advancing medical research and clinical interventions.

Mathematical and numerical modelling has become an indispensable approach to describing biophysical processes in the cardiovascular system \cite{quarteroni2019mathematical, Santiago2018FullySupercomputers, niederer2019computational, bucelli2023mathematical, zingaro2024electromechanics, caiazzo2012mathematical}. In the context of the heart, cardiomyocyte models are essential for characterising the flow of ions across cellular and intracellular membranes \cite{Mayourian2018AnArrhythmogenicity, 2024FamilialCardiomyopathies, Trayanova2024ComputationalTranslation}. The extension of these models to include gap junctions between adjacent cells enables the simulation of ionic currents in the myocardium, providing a comprehensive view of cardiac electrophysiology \cite{Mayourian2018AnArrhythmogenicity, 2024FamilialCardiomyopathies, Trayanova2024ComputationalTranslation}. Moreover, various mechanical models have been proposed to describe the motion of the heart, elucidating deformation patterns, stress distribution, and the mechanical properties of cardiac tissues \cite{hunter1998modelling, Nash2000ComputationalHeart, Regazzoni2018ActiveInteractions, Mythri2024First-PrinciplesHeart}. Recent advancements in multi-physics models have integrated electrophysiology and solid mechanics to accurately reflect the excitation-contraction coupling that occurs within the myocardium \cite{Regazzoni2022ACirculation, Augustin2016AnatomicallyDeformation, Lafortune2012CoupledFormulation, Trayanova2011ElectromechanicalVentricles, Strocchi2020SimulatingPericardium}. Regarding the vasculature, fluid dynamic models are used to simulate blood flow in vessels, providing insights into shear stress distribution, pressure gradients, and flow features \cite{Xiao2013Multi-scaleNetwork, Zavodszky2024CellularHemoCell, Wang2024AnDisease, McCullough2023HighWillis, Qureshi2014NumericalCirculation, Kamada2022BloodDiseases}. Various mechanical models have also been employed to analyse the behaviour of vessel walls \cite{Coccarelli2021ASimulations, Wang2022Image-BasedArteries, MasoTalou2018MechanicalStudies, Kalita2008MechanicalWalls}. Moreover, advancements in multi-physics models have coupled fluid dynamics and solid mechanics to capture the interplay between blood flow and wall motion \cite{Bazilevs2009Patient-specificDevice, Filonova2020VerificationSolution, Fringand2024AMethods, MansillaAlvarez2022AVessels, Syed2023ModelingMethods}. These models collectively enhance our understanding of cardiovascular physiology.

To fully characterise human cardiac function and blood transport, it is essential to consider the complex interactions between the heart and blood vessels. However, most existing models focus on one organ and make simplifying assumptions about the other. Typically, detailed cardiac function models use reduced-order models to represent blood circulation \cite{Regazzoni2022ACirculation, Augustin2021ACirculation, Gerach2021Electro-mechanicalApproach, Trayanova2011ElectromechanicalVentricles, fedele2023comprehensive}, while detailed vascular blood flow models often prescribe a profile of cardiac outflow, based on measurements, statistics, or theories \cite{Stokes2023AneurysmalIndices, Tajeddini2024TypeSimulation, Marzo2011ComputationalConditions, Kamada2022BloodDiseases}. Although these assumptions reduce computational costs, neglecting the feedback mechanism between the two subsystems is likely to impair the accuracy of the biophysical processes described. In contrast, a multi-component model, which integrates high-fidelity representations of the biophysical components, offers a more realistic and comprehensive description of the full system.

\begin{figure}[!htb]
    \centering
    \includegraphics[trim = {0cm 0cm 0cm 0cm}, clip, width=\textwidth]{figures/coupled-model.pdf}
    \caption{\rev{The coupled multi-component, multi-physics human cardiovascular model constructed in this work.}}
    \label{fig:coupled_model}
\end{figure}

In this paper, we present an innovative approach to couple a 3D electromechanical model of the heart with a 3D fluid mechanics model of \rev{vascular blood flow} (see Figure \ref{fig:coupled_model}). We adopt a multi-component coupling strategy rather than rewriting everything into a new, monolithic code, leveraging existing specialised solvers to build a more comprehensive model. To address this complexity, we employ a file-based partitioned coupling scheme, which couples the two models by alternately writing and reading data through intermediate files. This approach allows each model to run independently while sharing essential simulation data, thereby maintaining optimal computational performance.

We establish the validity of this approach by using Alya \cite{Vazquez2016Alya:Exascale, Santiago2018FullySupercomputers} to model cardiac electromechanics and HemeLB \cite{Mazzeo2008HemeLB:Geometries, Zacharoudiou2023DevelopmentSimulation} to model vascular blood flow. These models have been developed by separate research groups, emphasise \rev{disparate} dynamical scales, utilise distinct discretisation schemes, \rev{and are written in different programming languages}. By leveraging the excellent scaling performance of these models on supercomputers, this study aims to construct an efficient multi-component, multi-physics model of the cardiovascular system \rev{and investigate its potential for simulating realistic physiological and pathological conditions}. Moreover, it illustrates how a virtual human \cite{Hoekstra2018VirtualClinic}, also referred to as a human digital twin, can be developed by integrating efforts from different research groups working in complementary, specialised fields.

The subsequent sections of this paper are structured as follows. In Section \ref{sec:methods}, we present the individual and coupled models of cardiac function and blood flow, along with the simulation tests designed to verify the coupled model. In Section \ref{sec:results_discussion}, we analyse the results of \rev{four simulation tests designed to evaluate different aspects of the coupled model, including verification of the coupling scheme, quantification of coupling effects, and application to studying the impact of myocardial scarring on vascular flow}. In Section \ref{sec:limitations}, we discuss the limitations of the coupled model. Finally, we conclude our findings in Section \ref{sec:conclusion}.

%% file: 2_Methods.tex
\section{Methods} \label{sec:methods}
In this section, we first present the electromechanical model of the heart in Section \ref{sec:methods-alya} and the fluid mechanics model of vascular blood flow in Section \ref{sec:HemeLB}. Then, we describe the coupled model in Section \ref{sec:coupled-model} and the coupling scheme in Section \ref{sec:coupling_scheme}. Lastly, we outline the setup of simulation cases for verifying disparate aspects of the coupled model in Section \ref{sec:simulations}.

\subsection{Electromechanical Model of the Heart in Alya}
\label{sec:methods-alya}
To model the ventricular electrical activity and mechanical contraction, we use a fully-coupled electromechanical model developed in Alya \cite{Vazquez2016Alya:Exascale, Santiago2018FullySupercomputers}: a simulation code designed for solving multi-physics problems in engineering applications using \rev{the} finite element method. \rev{Written in Fortran, the} code is designed for high-performance computing and has demonstrated scalability up to 100,000 cores \cite{Vazquez2016Alya:Exascale}.

\begin{figure}[htb!]
\stackinset{l}{85pt}{t}{5pt}{\Large $\boxed{\domain}$}{%
\stackinset{l}{130pt}{t}{10pt}{\Large ${\gammabase}$}{%
\stackinset{l}{25pt}{t}{5pt}{\Large $\gammaendo$}{%
\stackinset{l}{110pt}{t}{160pt}{\Large $\gammaepi$}{%
\includegraphics[trim = {0cm 4cm 0cm 0cm}, clip, width=\textwidth]{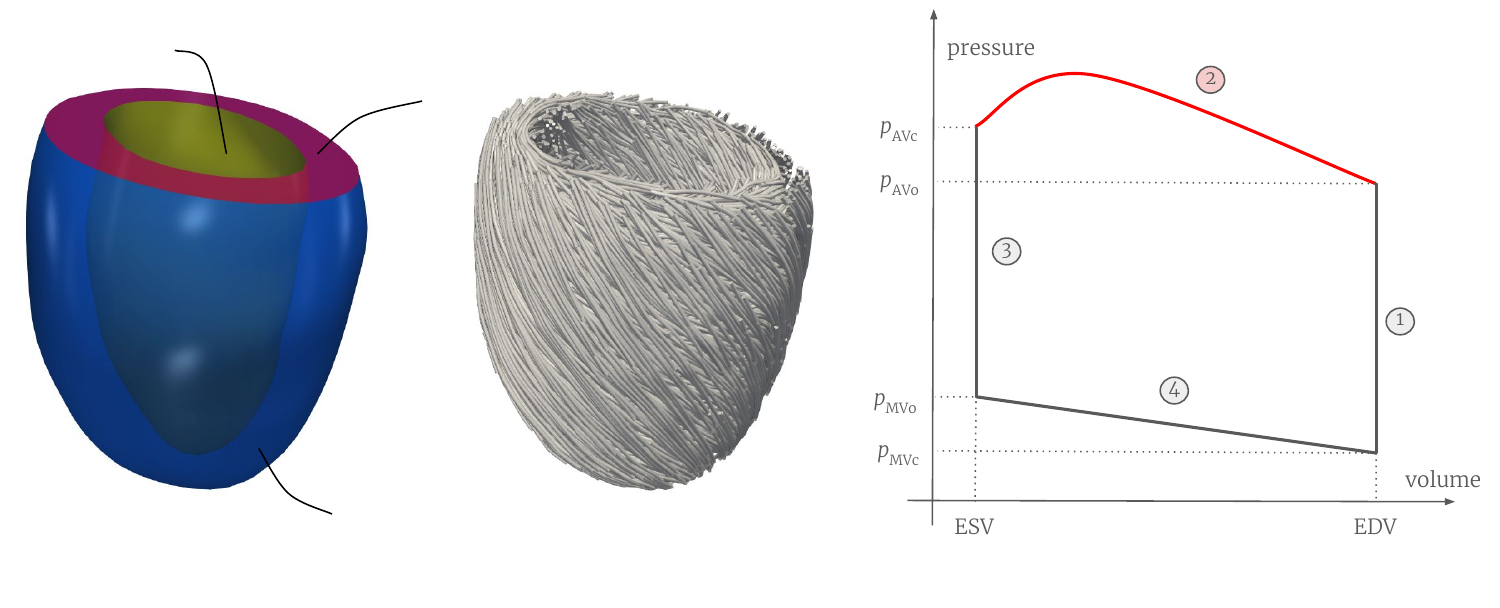}
}}}}
\caption{The electromechanical model in Alya. From left to right: left ventricular geometry with its boundary portions, cardiac fibres and pressure-volume loop. In the latter, we highlight in red the ejection phase (phase 2), i.e. where the coupling with HemeLB is carried out.}
\label{fig:lv-boundary-fibers-pvloop}
\end{figure}

Let \rev{$\domain \subset \R{3}$} be the end-diastolic configuration of a left ventricle at time $t=0$, also representing our \textit{reference configuration}, and $T$ the final time. As we sketch in Figure \ref{fig:lv-boundary-fibers-pvloop}, $\domain$ is bounded by $\partial \domain = \gammaendo \cup \gammaepi \cup \gammabase$, which are the endocardium, epicardium, and base respectively; we denote by $\normal$ the outward-pointing normal to the boundary.

We model cardiac electrophysiology via the monodomain equation \cite{franzone2014mathematical}. Let $v: \domain \times (0, T) \to \R{}$ be the transmembrane potential, $\gating: \domain \times (0, T) \to \R{n_{\gating}}$ the gating variables (with $n_{\gating}$ the number of gating variables), and $\concentration$ the ionic concentrations $\concentration: \domain \times (0, T) \to \R{n_{\concentration}}$ (with $n_{\concentration}$ the number of ionic concentrations inside of the cell). The monodomain model -- coupled to a cellular ionic model -- reads:
\begin{align}
        & \pdv{v}{t} - \div (\D \grad v ) + \frac{\Iion (v, \gating, \concentration)}{\Cm}  = \frac{\Iapp}{\Cm} & \text{ in } \domain \times (0, T),\label{eq:mono-parabolic-diff}
        \\
        & \frac{\mathrm d \gating }{\mathrm d t} = \bm W (v, \gating, \concentration ) & \text{ in } \domain \times (0, T) ,\label{eq:mono-gating}
        \\
        & \frac{\mathrm d \concentration }{\mathrm d t} = \bm C (v, \gating, \concentration ) & \text{ in } \domain \times (0, T) ,\label{eq:mono-concentration}
\end{align}
with homogeneous boundary conditions on the whole boundary. $\D$ is the orthotropic tensor of local diffusivity in the reference configuration expressed, with respect to the tensor of conductivity $\condtens$, as \cite{bueno2008minimal}:
\begin{equation*}
    \D = \frac{\condtens}{\Cm \Sv} = \frac{1}{\Cm \Sv} \left (
   \condf \fzero \otimes \fzero +
   \conds \szero \otimes \szero +
   \condn \nzero \otimes \nzero
    \right ),
\end{equation*}
where 
$\condf$, $\conds$, and $\condn$ are the conductivity in the fibre, sheet and normal directions and $\fzero$, $\szero$, and $\nzero$ the unit vector field expressing their directions. Cardiac fibres are modelled with Laplace-Dirichlet rule-based methods \cite{doste2019rule, piersanti2021fibers}, as depicted in Figure \ref{fig:lv-boundary-fibers-pvloop}. $\Cm$ is the membrane capacitance per unit area and $\Sv$ the surface-to-volume ratio. $\Iapp(\X, t)$ is the applied current; $\bm W$ and $\bm C$ are the right-hand sides of the system of ODEs corresponding to $\gating$ and $\concentration$, respectively. The term $\Iion$ collects the ionic currents. \rev{We use the cellular ionic model proposed by \citet{o2011simulation} and incorporate} the modifications proposed by \citet{passini2016mechanisms} and the modified conductance as described by \citet{dutta2017}.

Let $\deformationoperator: \domain \times (0, T) \to \R{3}$ be a deformation operator which maps each point $\X$ of the reference configuration $\domain$ onto the \textit{current configuration} $\domaint=\deformationoperator(\domain, t)$, for all $t\in (0, T)$, so that each point of the current configuration is $\x(\X, t) = \deformationoperator(\X, t)$. We define the displacement field $\displacement: \domain \times (0, T) \to \R{3}$ as $\displacement(\X, t) = \x(\X, t) - \X$, the deformation tensor  $F(\X, t) = \pdv{\deformationoperator}{\X}$, and $J$ is the determinant of $F$: $J=\det (F) > 0$.  
\rev{We consider an active stress formulation, for which we can decompose the second Piola-Kirchhoff stress tensor $S$ into a passive and an active term:
\begin{equation*}
    S(\displacement, \tact (\qact)) = \Spass(\displacement) +  \Sact(\tact(\qact)),
\end{equation*}
where $\tact$ is the active tension and $\qact$ a vector of state variables associated with the contractile mechanism in cardiomyocytes \cite{Levrero-Florencio2020SensitivityBiomarkers, land2017model}.
}
The mechanical deformation is modelled by Cauchy's first equation of motion equipped with initial and boundary conditions:

\begin{align}
    & \density^\mathcal{A} \pdv{^2\displacement}{t^2} - \div\left (\FS \right ) = \bm 0 & \text{ in } \domain \times (0, T), \label{eq:strong-form-pde}
    \\
    & \rev{\density J_0 = \rho^\mathcal{A} J} & \rev{\text{ in } \domain \times (0, T), \label{eq:strong-form-pde-mass}}
    \\
    & \FS \normal = (\kepi \displacement \cdot \normal) \normal & \text{ on } \gammaepi \times (0, T), \label{eq:strong-form-bcepi}
    \\
    & \displacement \cdot \hat{\bm e}_x = \displacement \cdot \hat{\bm e}_y = 0  &  \text{ on } \gammabase \times (0, T), \label{eq:strong-form-bcbase}
    \\
    & \FS \normal = - \pendo \JFmT \normal & \text{ on } \gammaendo \times (0, T), \label{eq:strong-form-bcendo}
    \\
    & \displacement = \bm 0 \text{ and } \FS = \bm 0 & \text{ in } \domain \times \{ 0\}, \label{eq:strong-form-ic}
\end{align}
where $\density^\mathcal{A}$ is the tissue density \rev{in the reference configuration}. \rev{To model $\Sact$, i.e. to model the excitation-contraction coupling, we use the active force generation model introduced by \citet{land2017model}. This model consists of a set of ODEs describing the local dynamics of the vector $\qact$:
\begin{equation*}
    \frac{\mathrm d \qact}{\mathrm d t} = \bm Q (\qact, \concentration, \lambdaf, \dot{\lambdaf}) \quad \text{ in } \domain \times (0, T),
\end{equation*}
where $\lambdaf$ is the stretch in the direction of the fibres, $\bm Q$ is the right-hand side of the set of ODEs and $\tact = \tact (\qact, \lambdaf)$. For additional details on the excitation-contraction model, we refer the readers to \citet{land2017model} and \citet{Levrero-Florencio2020SensitivityBiomarkers} for its specific implementation in Alya.

To model the passive part $\Spass$, we consider the Holzapfel and Ogden constitutive law \cite{Holzapfel2009ConstitutiveCharacterization}.}

We set a null initial condition in terms of both displacement and stress, as expressed in equation \ref{eq:strong-form-ic}. In equation \ref{eq:strong-form-bcepi}, we set Robin-type boundary conditions on the epicardium with $\kepi$ the corresponding normal spring stiffness. We orient the ventricular geometry such that the normal to the base coincides with the $z$ axis. On the base, we set the displacement to zero in the $x$ and $y$ direction, leaving a stress-free condition in the remaining direction $z$. In equation \ref{eq:strong-form-bcbase}, $\hat{\bm e}_x$ and $\hat{\bm e}_y$ are the unit vectors in the $x$ and $y$ directions, respectively. On the endocardium, as expressed in equation \ref{eq:strong-form-bcendo}, we set a Neumann boundary condition with the time-dependent ventricular blood pressure $\pendo(t)$. 

To model the \rev{ventricular blood pressure $\pendo(t)$}, we subdivide the heartbeat into four phases: isovolumetric contraction, ejection, isovolumetric relaxation, and filling (see Figure \ref{fig:lv-boundary-fibers-pvloop}). \rev{We denote by $V$ be the volume of the blood contained within the left ventricular cavity.} After an initialization phase that allows the cardiac muscle to be brought to the time of end-diastole, we model each phase of the heartbeat as follows \cite{Levrero-Florencio2020SensitivityBiomarkers}: 
\begin{enumerate}
    \item \textit{Isovolumetric contraction}: starting from end-diastole configuration, when the ventricular volume stops increasing ($\dert{}{V}\leq0$), the mitral valve closes, the ventricle contracts and the pressure increases keeping a constant volume equal to the end-diastolic volume (EDV). We integrate in time the following laws:
    \begin{equation*}
        \frac{\mathrm d \pendo(t)}{\mathrm d t} = -c_1 \frac{\mathrm d V(t)}{\mathrm d t} - c_2 \frac{\mathrm{d}^2V(t)}{\mathrm{d}t^2},
    \end{equation*}
    where $c_1$ and $c_2$ are two coefficients that allow us to keep the condition of constant volume. The second term, acting on the second-order volume derivative, stabilizes potential spurious oscillations in the pressure \cite{Levrero-Florencio2020SensitivityBiomarkers}. \rev{The coefficients $c_1$ and $c_2$ are selected \textit{a posteriori} to ensure that these two conditions are met.}
    \item \textit{Ejection} starts when the ventricular pressure overcomes the systemic arterial pressure: $\pendo \geq \pavo$, where $\pavo$ is prescribed. The aortic valve opens, the muscle keeps contracting and the ventricular volume decreases. This dynamics is modelled using a two-element Windkessel (WK2) model \cite{Westerhof2009TheWindkessel, Shi2011ReviewSystem}:
    \begin{equation*}
        C^\mathcal{A} \dert{}{\pendo(t)} + \frac{1}{R^\mathcal{A}}\pendo(t) = - \dert{}{V(t)},
    \end{equation*}
    where $R^\mathcal{A}$ represents the total resistance of the arterial network and $C^\mathcal{A}$ the compliance.
    \item \textit{Isovolumetric relaxation} starts when the ventricular volume stops decreasing: $\dert{}{V(t)}\geq 0$: the aortic valve closes ($\pendo = \pavc$), the ventricle relaxes, the pressure decreases at a constant volume equal to the end-systolic volume (ESV). We integrate in time the following law:
    \begin{equation*}
        \frac{\mathrm d \pendo(t)}{\mathrm d t} = -r_1 \frac{\mathrm d V(t)}{\mathrm d t} - r_2 \frac{\mathrm{d}^2V(t)}{\mathrm{d}t^2}.
    \end{equation*}
    \rev{In analogy with the isovolumetric relaxation phase, }$r_1$ and $r_2$ are two coefficients that allow us to keep the volume constant \rev{ and to avoid spurious pressure oscillation, respectively. Both parameters are selected \textit{a posteriori} to ensure that both conditions are properly met.}
    \item \textit{Filling}: when the ventricular pressure reaches the atrial pressure $\pendo \leq \pmvo$ (where $\pmvo$ is prescribed),  the mitral valve opens, and the filling phase (from the atrium to the ventricle) starts. The ventricular pressure is modelled with the following decay equation:
    \begin{equation*}
        \dert{}{\pendo(t)} = - \gamma \dert{}{V(t)},
    \end{equation*}
    where $\gamma$ is a decay constant.
\end{enumerate}
In our baseline electromechanical simulation (Test 0), i.e. using Alya in a standalone fashion, we model the endocardial pressure as explained above. When we couple Alya with HemeLB, as we shall discuss in Section \ref{sec:coupled-model}, the endocardial pressure is instead given by HemeLB in each time step. 

\rev{In Alya, the volume $V$ of the blood contained within a cavity can be computed using two different methods. The first method applies the divergence theorem, as explained in \citet{Levrero-Florencio2020SensitivityBiomarkers}, while the second method relies on the mixed-product approach. In the latter, the open surface $\gammaendo$ is closed with a plane $\gammaplane$, which is then triangulated. This process forms a closed surface $\gammapool = \gammaendo \cup \gammaplane$, with $\gammaendo \cap \gammaplane = \emptyset$. Let $N_\mathrm{e}$ be the total number of elements (2D triangles) in the closed surface $\gammapool$, and let $\bm{a}_i$, $\bm{b}_i$ and $\bm{c}_i$ be the position vectors in the current configuration (dependent on $\displacement$) for each vertex of each triangle, with $i = 1, \dots, N_\mathrm{e}$. The volume enclosed by $\gammapool$ is then given by
\begin{equation}
    V(\displacement) = \frac{1}{6} \sum_{i=1}^{N_\mathrm{e}} \bm a_i (\displacement) \cdot (\bm b_i (\displacement) \times \bm c_i (\displacement)).
    \label{eq:volume-alya}
\end{equation}
Notice that this volume depends on the displacement $\displacement$. According to our experience, this method provides a faster computation of the volume compared to the one using the divergence theorem.
} 

Spatial discretisation \rev{of the partial differential equations} is carried out with linear finite elements. We use a first-order Yanenko operator splitting scheme and the implicit Euler time advancing scheme for electrophysiology (see \citet{Santiago2018FullySupercomputers} for additional details). We solve the non-linear mechanical problem with the Newton-Raphson scheme. Time discretisation of the mechanical problem is carried out with the Newmark scheme using the Rayleigh-damping method to damp potential spurious oscillations \cite{belytschko2014nonlinear, Santiago2018FullySupercomputers}.

\subsection{Fluid Mechanics Model of Vascular Blood Flow in HemeLB} \label{sec:HemeLB}
The 3D blood flow in vessels is simulated using HemeLB \cite{Mazzeo2008HemeLB:Geometries, HemeLB}, an open-source fluid flow solver based on the lattice Boltzmann method (LBM) \cite{Succi2018TheMatter, Kruger2017TheMethod}. This solver was verified and validated in previous works examining vascular flows \cite{Nash2014ChoiceDomains, Groen2018ValidationMeasurements, McCullough2021HighFistula}. HemeLB has been developed into versions optimised for CPU and GPU architectures. \rev{Written in C++, both} versions have demonstrated excellent scaling characteristics on high-performance computers: the CPU version has utilised over 300,000 CPU cores, while the GPU version has leveraged over 18,000 GPUs \cite{Zacharoudiou2023DevelopmentSimulation}. In this study, we use the CPU version of HemeLB \cite{HemePure}.

Established on the basis of kinetic theory, the LBM solves for the velocity distribution function $f$ of fluid particles at the mesoscopic scale in the Boltzmann equation rather than the fluid density $\rho$, flow velocity $\bm{U}$, and fluid pressure $P$ at the macroscopic scale \cite{Succi2018TheMatter, Kruger2017TheMethod}. Discretisation of the Boltzmann equation over an isotropic lattice gives the lattice Boltzmann equation
\begin{equation*}
    f_i (t + \Delta t^\mathcal{H}, \bm{x} + \bm{c_i} \Delta t^\mathcal{H}) = f_i (t, \bm{x}) + \Omega_i (t, \bm{x}) \Delta t^\mathcal{H}.
\end{equation*}
Here, $\Omega_i$ is the collision operator which characterises the interactions occurring between particles, and $\{f_i, i=0,\dots,q\}$ are the discretised form of $f$ in $q$ different velocities $\{\bm{c_i}\}$ \cite{Succi2018TheMatter, Kruger2017TheMethod}.

In this work, we use the Bhatnagar-Gross-Krook \cite{Bhatnagar1954ASystems} collision operator $\Omega_i = - (f_i - f^{eq}_i) / \tau_0$, which relaxes the distribution functions towards an equilibrium $f^{eq}_i$ with a relaxation time $\tau_0$, and the D3Q19 velocity set (3D lattice, $q=19$) \cite{Kruger2017TheMethod}. The lattice Boltzmann equation with the Bhatnagar-Gross-Krook collision operator, known as the LBGK equation, reads \cite{Bhatnagar1954ASystems, Kruger2017TheMethod, Succi2018TheMatter}
\begin{equation}
    f_i (t + \Delta t^\mathcal{H}, \bm{x} + \bm{c_i} \Delta t^\mathcal{H}) = f_i (t, \bm{x}) - \frac{\Delta t^\mathcal{H}}{\tau_0} \Big[ f_i (t, \bm{x}) - f^{eq}_i (t, \bm{x}) \Big].
    \label{eq:lb}
\end{equation}
This equation corresponds to the incompressible Navier-Stokes equations
\begin{subequations}
    \begin{align}
        \nabla \cdot \bm{U} &= 0, \label{eq:NSE_continuity} \\
        \frac{\partial \bm{U}}{\partial t} + (\bm{U} \cdot \nabla) \bm{U} &= - \frac{1}{\rho} \nabla P + \nabla \cdot (\nu_0 \nabla \bm{U}), \label{eq:NSE_momentum}
    \end{align}
\end{subequations}
with a \rev{kinematic} viscosity $\nu_0$ given by
\begin{equation} \label{eq:kinematic_viscosity}
    \nu_0 = c_s^2 \bigg( \tau_0 - \frac{\Delta t^\mathcal{H}}{2} \bigg),
\end{equation}
when the Mach number
\begin{equation} \label{eq:Mach}
    \mathit{Ma} = \lVert \bm{U} \rVert / c_s
\end{equation}
is sufficiently low ($\ll 1$), where $c_s = (1 / \sqrt{3}) \Delta x^\mathcal{H} / \Delta t^\mathcal{H}$ is the speed of sound \cite{Succi2018TheMatter, Kruger2017TheMethod}. Using the distribution functions, the macroscopic variables can be calculated as
\begin{align}
    \rho &= \sum_{i=0}^{q} f_i, \label{eq:lb-density}\\
    \bm{U} &= \frac{1}{\rho} \sum_{i=0}^{q} f_i \bm{c_i}, \label{eq:lb-velocity}\\
    P &= \rho c_s^2. \label{eq:lb-pressure}
\end{align}
In general, the Mach number is non-zero and the Navier-Stokes equations reproduced by the LBM are weakly compressible. The compressibility errors associated with the LBGK model typically scale with $\mathcal{O}(\mathit{Ma}^2)$. Because of these errors, the density fluctuates around an average value $\rho_0^\mathcal{H}$, which corresponds to the reference pressure $P_\text{ref}$ of the fluid:
\begin{equation} \label{eq:lb-pressure_ref}
    P_\text{ref} = \rho_0^\mathcal{H} c_s^2.
\end{equation}

\begin{figure}[htb!]
\centering
\stackinset{l}{110pt}{t}{200pt}{\Large $\boxed{\domainheme}$}{%
\stackinset{l}{25pt}{t}{190pt}{\Large $\gammainheme$}{%
\stackinset{l}{0pt}{t}{65pt}{\Large $\gammawallheme$}{%
\stackinset{l}{125pt}{t}{20pt}{\Large ${\gammaoutheme}_{,0}$}{%
\stackinset{l}{80pt}{t}{10pt}{\Large ${\gammaoutheme}_{,1}$}{%
\stackinset{l}{35pt}{t}{-20pt}{\Large ${\gammaoutheme}_{,2}$}{%
\stackinset{l}{0pt}{t}{280pt}{\Large ${\gammaoutheme}_{,3}$}{%
\includegraphics[width=0.3\textwidth]{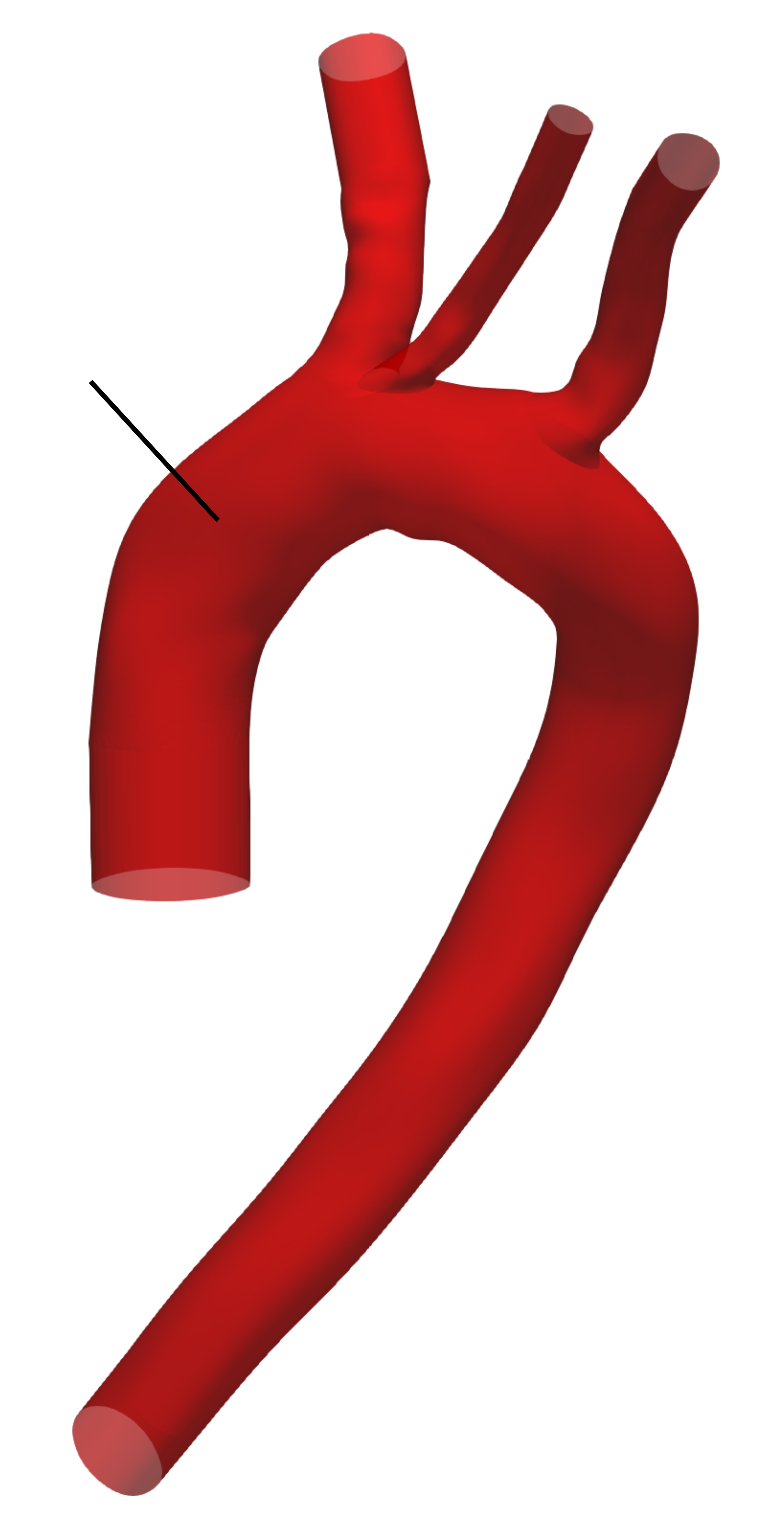}
}}}}}}}
\caption{The thoracic aorta model for HemeLB in Test 2. It is adapted from the 0012\_H\_AO\_H model in the Vascular Model Repository \cite{Wilson2013TheResults} produced from a computed tomography angiogram of a 26-year-old male human. The fluid domain $\domainheme$ is the interior of this model bounded by the vessel wall $\gammawallheme$, an inlet $\gammainheme$, and four outlets $\{{\gammaoutheme}_{,i}, i = 0,1,2,3\}$.}
\label{fig:thoracic_aorta}
\end{figure}

Let \rev{$\domainheme \subset \R{3}$} be the fluid domain of a vascular model. It is bounded by the vessel wall $\gammawallheme$, an inlet $\gammainheme$, and a group of outlets $\gammaoutheme = \{{\gammaoutheme}_{,i}\}$. As an illustration, the arterial model used in this study is shown in Figure \ref{fig:thoracic_aorta}.

In the fluid domain, $\domainheme$, the blood is assumed to be Newtonian with a constant dynamic viscosity of 0.0035 Pa s \cite{Kannojiya2021SimulationAspects} and a constant density ($\rho_0^\mathcal{H}$) of 1050 kg m$^{-3}$ \cite{Thomas2016BloodReview}. \rev{This results in a kinematic viscosity ($\nu_0$) of $3.33 \cdot 10^{-6}$ m$^2$/s.} The vessel walls, $\gammawallheme$, are assumed to be rigid by imposing the Bouzidi-Firdaouss-Lallemand boundary condition \cite{Bouzidi2001MomentumBoundaries} on the walls, $\gammawallheme$.

The inlet section, $\gammainheme$, serves as the coupling interface in all simulations conducted in this study. The inlet flow rate $\Qin$, provided by Alya through coupling, is used to calculate the flow velocity at each lattice site on the inlet plane, as described in our previous work \cite{Lo2022ParametricNetworks}. The velocity is assumed to be perpendicular to the inlet plane and follows a parabolic profile:
\begin{equation} \label{eq:inlet_velocity}
    \bm{U}(r,t) = \frac{g(r)}{\int_{\gammainheme} g(r) \ \mathrm{d} A} \ \Qin(t) \ \vu{n},
\end{equation}
where $\vu{n}$ is the unit normal vector, $A$ denotes the area, and
\begin{equation} \label{eq:parabolic_profile}
    g(r) = 1 - \frac{r^2}{r_{\max}^2}
\end{equation}
is a function of the distance $r$ between the lattice site and the inlet centre. On the edge of the inlet plane where $r = r_{\max}$, this function equals zero to satisfy the no-slip boundary condition. The speed is then imposed as an inlet condition using Ladd's approach \cite{Ladd1994NumericalFoundation}.

At the outlet sections, $\gammaoutheme$, the pressure boundary condition \rev{outlined in \citet{Nash2014ChoiceDomains}} is imposed. The pressure $P$ is calculated from the flow rate $Q$ using \rev{a three-element Windkessel (WK3)} model \cite{Westerhof2009TheWindkessel, Shi2011ReviewSystem, Capoccia2015DevelopmentAssistance}, where the flow rate is obtained by integrating the flow velocity in the normal direction over the outlet plane. \rev{The WK3 model extends the WK2 model outlined in \citet{Lo2022ParametricNetworks} by introducing an additional resistance in series with the WK2 components. This extra resistance, known as the characteristic impedance $R_c$, accounts for the flow resistance of the proximal vessels and improves the accuracy of the captured flow waveforms, particularly in the medium-to-high-frequency range. In addition, spurious oscillations can be avoided by matching the characteristic impedance with the flow resistance of the proximal vessels \cite{Formaggia2006NumericalHeart}. The governing equation of the WK3 model is
\begin{equation} \label{eq:WK3_ODE}
    R_p C_p \frac{\mathrm{d} P}{\mathrm{d} t} + P = R_c R_p C_p \frac{\mathrm{d} Q}{\mathrm{d} t} + (R_c + R_p) Q,
\end{equation}
where $R_p$ and $C_p$ are components of the WK2 model representing the resistance and capacitance of peripheral vessels, respectively. Following the approach of \citet{Grinberg2008OutflowOutlets}, we discretise equation \ref{eq:WK3_ODE} using a semi-explicit scheme with first-order approximations. The pressure at time step $n + 1$ is given by
\begin{equation}
    P^{[n+1]} = \frac{R_p C_p}{R_p C_p + \Delta t^\mathcal{H}} P^{[n]} + \frac{(R_c + R_p) \Delta t^\mathcal{H}}{R_p C_p + \Delta t^\mathcal{H}} Q^{[n]} + \frac{R_c R_p C_p}{R_p C_p + \Delta t^\mathcal{H}} \Big( Q^{[n]} - Q^{[n-1]} \Big).
\end{equation}
The choice of the parameters $\{R_c, R_p, C_p \}$ requires problem-specific tuning. The values used for our tests are provided individually in Sections \ref{sec:test1} and \ref{sec:test2}.}

\begin{figure}[!htb]
    \centering
    \includegraphics[width=0.6\linewidth]{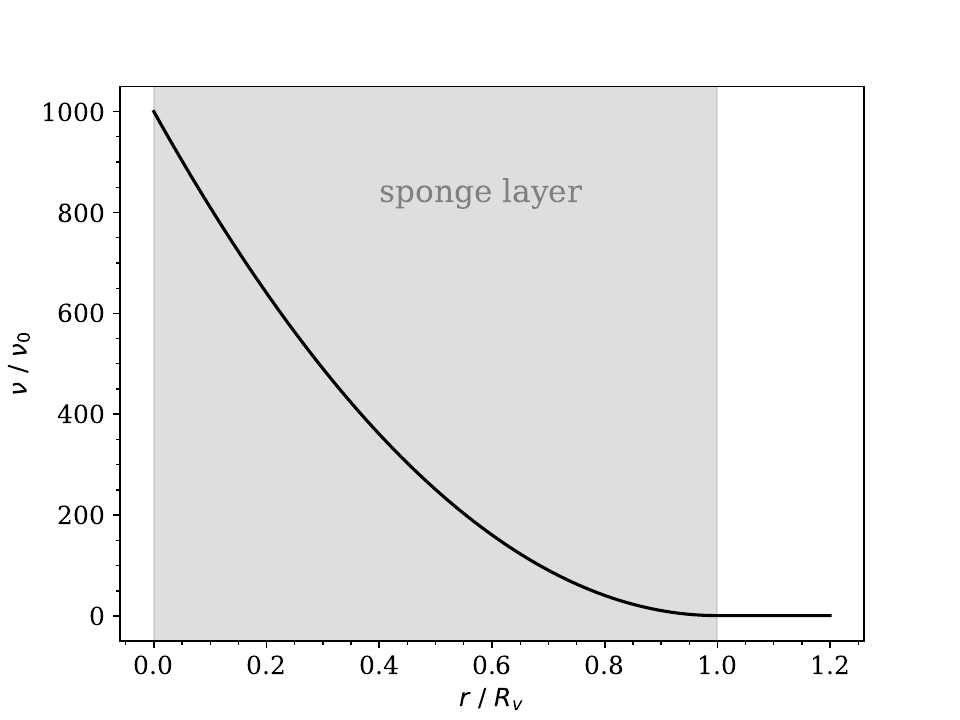}
    \caption{Viscosity $\nu$ within a viscous sponge layer as a function of the distance from the outlet centre $r$, normalised by the base \rev{kinematic} viscosity $\nu_0$ and the radius of the sponge layer $R_v$ respectively. The amplification factor of the viscosity $\nu/\nu_0$ smoothly increases from 1 on the sponge layer surface ($r=R_v$) to 1000 at the outlet centre ($r=0$).}
    \label{fig:sponge_layer_viscosity}
\end{figure}

We include viscous sponge layers around the outlets to prevent reflections from the boundaries \cite{Vergnault2012AFlow}. Each outlet is surrounded by a spherical sponge layer with a radius $R_v$ at its midpoint. Within the sponge layer ($r \leq R_v$), the \rev{kinematic} viscosity $\nu$ gradually increases from the base value $\nu_0$ at the spherical surface ($r=R_v$) to the maximum value $\nu_{\max}$ at the centre ($r=0$). This variation is described by a function of the distance from the outlet centre $r$ as
\begin{equation}
    \nu(r) = \nu_0 + (\nu_{\max} - \nu_0) \Big( \frac{r}{R_v} - 1 \Big)^2
\end{equation} and plotted in Figure \ref{fig:sponge_layer_viscosity}. In this study, the ratio $\nu_{\max}/\nu_0$ is set to 1000, as in other studies \cite{Vergnault2012AFlow, Ezzatneshan2019ImplementationFlows, Xue2024TheAorta}. The change in viscosity is implemented by adjusting the relaxation time, according to their relationship given in equation \ref{eq:kinematic_viscosity}, as follows:
\begin{equation}
    \tau(r) = \frac{1}{2} + \frac{\nu(r)}{c_s^2}.
\end{equation}
Outside the sponge layer ($r > R_v$), both the viscosity and the relaxation time remain at their base values $\nu_0$ and $\tau_0$ respectively.

In an incompressible flow, the gradient rather than the value of the fluid pressure is what matters, as evidenced in equation \ref{eq:NSE_momentum}; however, in this coupled system involving mechanics, the pressure value itself becomes important due to its physical role. To ascertain this value within the fluid flow, the reference pressure, $P_\text{ref}$, is set in HemeLB. This reference is chosen to be close to the pressure on the endocardium before the coupling starts to prevent large oscillations in the solutions.

\subsection{Coupled Model}
\label{sec:coupled-model}
We couple the two models during the ejection phase only (corresponding to phase 2 mentioned in Section \ref{sec:methods-alya}). Thus, for the coupling, we restrict the time domain to $\left [ t_0, t_f \right ]$, where $t_0$ and $t_f$ are the initial and final time instants of the ejection phase. Before the coupling begins, one heartbeat is simulated as a warm-up. Similarly, the flow velocity is increased steadily from zero to the value just before the coupling phase starts.

\begin{subequations}
To express the whole coupled model at a continuum level, we first compactly denote the ventricular electromechanical model --- solved in Alya --- as
\begin{equation} \label{eq:coupled-alya}
    \mathcal A (v, \gating, \concentration, \displacement; \pendo) = 0 \quad \text{in } \domain \times [t_0, t_f],
\end{equation}
consisting of the gathering of equations \ref{eq:mono-parabolic-diff}--\ref{eq:mono-concentration} and \ref{eq:strong-form-pde}--\ref{eq:strong-form-ic}. Similarly, we compactly denote the vascular hemodynamics model --- solved in HemeLB --- as
\begin{equation} \label{eq:coupled-heme}
    \mathcal H (\bm U, P ; \Qin) = 0 \quad \text{in } \domainheme \times [t_0, t_f],
\end{equation}
composed of equations \ref{eq:lb}, \ref{eq:lb-velocity} and \ref{eq:lb-pressure}.

The two models are coupled by enforcing the dynamic balance and kinematic continuity. For the continuity of stresses, we enforce that the endocardial ventricular pressure $\pendo$ on $\gammaendo$ equals the average pressure at the inlet section $\gammainheme$. This dynamic balance is formulated as
\begin{equation} \label{eq:coupled-dynamic}
    \pendo(t) = \frac{1}{\gammainheme} \int_{\gammainheme} P(\bm x, t) \, \mathrm{d} x  \quad \text{in } \gammaendo \times [t_0, t_f].
\end{equation}
Analogously, velocity continuity is enforced by assuming that the decreased rate of ventricular volume \rev{$V$} equals the inlet flow rate $\Qin$ across $\gammainheme$. This kinematic continuity is formulated as
\begin{equation} \label{eq:coupled-kinematic}
    \Qin(t) = - \frac{\mathrm d \rev{V}(\displacement)}{\mathrm d t} \quad \text{in } \gammainheme \times [t_0, t_f].
\end{equation}
\end{subequations}
\rev{The flow rate is computed using backward finite differences with the volume in equation \ref{eq:volume-alya}}.

Altogether, equations \ref{eq:coupled-alya} -- \ref{eq:coupled-kinematic} represent the fully coupled model at the continuum level.

A low-pass filter is implemented on the coupling interface to attenuate the spurious oscillations in the flow when needed. The flow rate value $\Qin$ given by Alya is passed through a simple RC circuit and results in the flow rate value $\widetilde{Q}_\mathrm{in}$ imposed at the inlet of the HemeLB domain. The relation between these flow rate values is governed by
\begin{equation}
    \Qin(t) - \widetilde{Q}_\mathrm{in}(t) = RC \frac{\mathrm{d} \widetilde{Q}_\mathrm{in}(t)}{\mathrm{d} t},
\end{equation}
where $R$ and $C$ are the resistance and capacitance of the RC circuit. The flow rate imposed at the inlet, $\widetilde{Q}_\mathrm{in}$, remains constant until $\Qin$ is updated by Alya. By discretising this equation in time and rearranging the terms, we obtain
\begin{equation} \label{eq:low-pass_filter}
    \widetilde{Q}_\mathrm{in}^{[n+1]} = \alpha \, \Qin^{[n+1]} + (1 - \alpha) \, \widetilde{Q}_\mathrm{in}^{[n]},
\end{equation}
where $\alpha = \Delta t^{\mathcal{A}} / (RC + \Delta t^{\mathcal{A}})$ is a smoothing factor lying within the range $[0, 1]$.

\subsection{Partitioned Coupling Scheme between the Two Models} \label{sec:coupling_scheme}
Alya and HemeLB are coupled via a partitioned scheme, wherein their respective model equations are solved independently using exchanged data. During intervals between data exchanges, Alya and HemeLB independently solve their model equations \ref{eq:coupled-alya} and \ref{eq:coupled-heme} using the exchanged data $\pendo$ and $\Qin$ as the boundary values on the coupling interfaces, respectively. In addition, Alya computes the flow rate $\Qin$ on $\gammaendo$ and transmits it to HemeLB, while HemeLB determines the pressure $\pendo$ on $\gammainheme$ and shares it with Alya.

The data exchange is achieved by file writing and reading mechanisms. Specifically, two intermediate files, File A and File B, are created in the working directory. File A is used to exchange $\pendo$, while File B is used for $\Qin$. This approach requires minimal changes in both codes, allowing for swift implementation.

\rev{Since Alya and HemeLB focus on different dynamic scales, they require different time step sizes. To accommodate this difference, the coupling scheme allows data exchange at different frequencies in the two solvers, controlled by the parameter $k$. This parameter is defined as the ratio of the time step size of Alya to that of HemeLB, i.e. $k = \Delta t^{\mathcal{A}} / \Delta t^{\mathcal{H}}$, and must be an integer greater than or equal to one. This constraint ensures that Alya, which models electromechanics on a larger dynamic scale, updates its boundary conditions at every time step, while HemeLB, operating at a mesoscopic fluid dynamics scale, updates less frequently. To manage data exchange, an iteration index $m \in [0, k-1]$ is introduced. Each time HemeLB completes a time step, $m$ is incremented by one. When $m$ reaches $k$, HemeLB retrieves the updated value of $\Qin$ from File B and resets $m$ to zero. This ensures that data exchange occurs once per time step of Alya and once every $k$ time steps of HemeLB.}

The algorithm for this coupling scheme is outlined in Algorithm \ref{alg:HeartCoupling_coupling}.

\begin{algorithm}[!htb]
	\caption{Algorithm for coupling Alya with HemeLB during the ejection phase.}
	\label{alg:HeartCoupling_coupling}
	\begin{algorithmic}[1]
		\State Initialise $\pendo^{[0]}$ and $\Qin^{[0]}$ at time step $n=0$.
	    \While {$n < (t_f - t_0) / \Delta t^{\mathcal{A}}$}
     	\State Alya reads $\pendo^{[n]}$ from File A.
	    \State Alya solves heart electromechanics (eq. \ref{eq:coupled-alya}) at time step $n$ using $\pendo^{[n]}$ for the endocardial boundary condition (eq. \ref{eq:strong-form-bcendo}).
	    \State Alya calculates $\Qin^{[n]}$ using eq. \ref{eq:coupled-kinematic}.
	    \State Alya writes $\Qin^{[n]}$ to File B
        \State Alya waits for $\pendo^{[n+1]}$ to be available in File A.
        \State HemeLB reads $\Qin^{[n]}$ from File B.
        \State Calculate $\widetilde{Q}_\mathrm{in}^{[n]}$ using eq. \ref{eq:low-pass_filter}.
        \State Initialise $m \leftarrow 0$.
        \While {$m < k$}
	    \State HemeLB solves vascular haemodynamics (eq. \ref{eq:coupled-heme}) at time step $n+(m+1)/k$ using $\widetilde{Q}_\mathrm{in}^{[n]}$ for the inlet boundary condition (eq. \ref{eq:inlet_velocity}).
        \State HemeLB calculates $\pendo^{[n+(m+1)/k]}$ using eq. \ref{eq:coupled-dynamic}.
        \State $m \leftarrow m+1$.
        \EndWhile
        \State HemeLB writes $\pendo^{[n+1]}$ to File A.
        \State HemeLB waits for $\Qin^{[n+1]}$ to be available in File B.
		\State $n \leftarrow n+1$.
        \EndWhile
	\end{algorithmic}
\end{algorithm}

\rev{While implicit coupling with sub-iteration is a viable option to improve the stability of the coupled model, it comes with a significant increase in computational cost. Given the novelty of our 3D multi-component, multi-physics cardiovascular model, our focus in this study is to establish its feasibility and demonstrate its potential for medical applications before exploring additional coupling techniques. In this study, we employ an explicit coupling scheme for its computational efficiency and simplicity.}

\subsection{Simulation Tests} \label{sec:simulations}
We conduct \rev{four} simulation tests to study different aspects of the proposed coupled model. Test 0 sets the baseline solution \rev{for both cardiac electromechanics and vascular haemodynamics} for subsequent comparisons with the coupled model. Test 1 verifies the coupling scheme using simplified geometries. Test 2 demonstrates the feasibility of the coupled model within physiological contexts \rev{and highlights its differences from the standalone simulations in Test 0. Test 3 applies the coupled model to study how myocardial scarring affects downstream vascular flow.}

\subsubsection{Test 0: \rev{Baseline Solutions for Left Ventricle and Thoracic Aorta}}
In this test, we perform \rev{standalone simulations using Alya and HemeLB separately.} These simulations establish the baseline solutions for comparisons with the coupled Alya-HemeLB simulations in Test 2. In addition, the results of the Alya simulation determine the reference pressure $P_\text{ref}$ in the fluid domain of HemeLB (see equation \ref{eq:lb-pressure_ref}).

\begin{table}[!htb]
\centering
\caption{Parameters and their corresponding descriptions, unit measures, and values for different physics models used in Alya to carry out the electromechanical simulations. For additional details on the parameters used in the electromechanical model, we refer readers to ref. \cite{Levrero-Florencio2020SensitivityBiomarkers}.}
\label{tab:alya-parameters}
\begin{tabular}{|l|c|p{5cm}|c|c|}
\hline
\textbf{Physics} & \textbf{Parameter} & \textbf{Description} & \textbf{Unit} & \textbf{Value} \\
\hline
Electrophysiology & $\bm{D}$ & Diffusivity in the fibers, normal and sheet directions & cm$^2$/ms & 0.03, 0.01, 0.01 \\
                  & $I_\mathrm{app}$ & Applied current & \textmu A/cm$^2$ & -40 \\
                  & $C_\mathrm{m}$ & Membrane capacitance per unit area & \textmu F/cm$^2$ & 1 \\
\hline
Haemodynamics     & $c_1$ & First coefficient for isovolumetric contraction & barye/ml & 0.05 \\
                  & $c_2$ & Second coefficient for isovolumetric contraction & barye s / ml & 10 \\
                  & $C^\mathcal{A}$ & Compliance of Windkessel model for ejection & ml/barye & \rev{0.0005} \\
                  & $R^\mathcal{A}$ & Resistance of Windkessel model for ejection & barye s / ml & \rev{2700} \\
                  & $r_1$ & First coefficient for isovolumetric relaxation & barye/ml & 0.1 \\
                  & $r_2$ & Second coefficient for isovolumetric relaxation & barye s / ml & 1 \\
                  & $\gamma$ & Decay constant for ventricular filling & barye / ml & -500 \\
                  & $\pavo$ & opening pressure of aortic valve & barye & 11998.8 \\
                  & $\pmvo$ & opening pressure of mitral valve & barye & 8000 \\
\hline
\end{tabular}
\end{table}

The geometry for \rev{the Alya} simulation represents the left ventricle, which is based on a 3D model of the human heart produced from computed tomography and magnetic resonance imaging scans and created by \citet{Zygote}. The endocardial pressure during the ejection phase is determined by the WK2 model, as detailed in Section \ref{sec:methods-alya}. The time step size $\Delta t^\mathcal{A}$ is set to $10^{-4}$ s for both electrophysiology and mechanics. The parameters of the physics models used in Alya are given in Table \ref{tab:alya-parameters}.

For the simulation performed with HemeLB, we employ a model of the thoracic aorta. This arterial model, shown in Figure \ref{fig:thoracic_aorta}, is adapted from the 0012\_H\_AO\_H model in the Vascular Model Repository \cite{Wilson2013TheResults}, which was produced from a computed tomography angiogram of a 26-year-old male human. The model includes the ascending aorta, the descending aorta, and three branches: brachiocephalic, left common carotid, and left subclavian arteries. The original model is scaled up by 1.25 times in all dimensions to ensure the inlet radius approximates the normal size for adults (0.015 m \cite{Hager2002DiametersTomography, Erbel2001DiagnosisCardiology}). In addition, the inlet and outlets are elongated by one to two times their diameters to mitigate instabilities caused by the boundary conditions \cite{Xiong2011SimulationProperties}.

The fluid domain for HemeLB is created by voxelising \cite{voxeliser} the thoracic aorta model to obtain a three-dimensional uniform grid of lattice sites. A voxel size of $\Delta x^\mathcal{H} = 10^{-4}$ m is used, resulting in about 150 lattice sites along a radius of the inlet \rev{and 253 million lattice sites in the entire simulation domain}. The time step size $\Delta t^\mathcal{H}$ is set to $6.25 \cdot 10^{-6}$ s, or $\tau_0 / \Delta t^{\mathcal{H}} = 0.50625$. This corresponds to a coupling frequency of $k = 16$ given the time step size of $10^{-4}$ s for Alya when the models are coupled in Test 2. For a physiological cardiac output resulting in a maximum flow rate $\Qin$ of around \rev{0.45} dm$^3$/s \cite{Hammermeister1974TheDisease}, the maximum velocity assuming a quadratic profile is \rev{1.27} m/s. The corresponding Mach number of \rev{0.138} is considered to contribute to a small compressibility error \cite{Kruger2017TheMethod}.

\begin{table}[!htb]
\centering
\caption{Parameters of the \rev{three}-element Windkessel model imposed at the outlets of the thoracic aorta model.}
\label{tab:WK2_params}
\begin{tabular}{|l|l|l|l|l|}
\hline
 & \textbf{Outlet 0} & \textbf{Outlet 1} & \textbf{Outlet 2} & \textbf{Outlet 3} \\ \hline
\rev{$R_c$ (kg m$^{-4}$ s$^{-1}$)} & \rev{$2.050 \cdot 10^{8}$} & \rev{$2.500 \cdot 10^{8}$} & \rev{$7.050 \cdot 10^7$} & \rev{$3.040 \cdot 10^{7}$} \\ \hline
\rev{$R_p$} (kg m$^{-4}$ s$^{-1}$) & $3.422 \cdot 10^{9}$ & $4.182 \cdot 10^{9}$ & $1.176 \cdot 10^9$ & $2.543 \cdot 10^{8}$ \\ \hline
\rev{$C_p$} (m$^4$ s$^2$ kg$^{-1}$) & $3.443 \cdot 10^{-10}$ & $2.817 \cdot 10^{-10}$ & $1.002 \cdot 10^{-9}$ & $4.632 \cdot 10^{-9}$ \\ \hline
\end{tabular}
\end{table}

The low-pass filter with an appropriate smoothing factor $\alpha$ as described in equation \ref{eq:low-pass_filter} is applied to stabilise the simulation. \rev{To determine the parameters of the WK3 models at the outlets, we first estimate $R_p$ and $C_p$ using the method we introduced previously \cite{Lo2022ParametricNetworks, Lo2024UncertaintySimulations} with the flow rate ratios $5.5:4.5:16:74$ for Outlets 0 to 3, respectively \cite{Boccadifuoco2018ValidationAnalysis}. After that, we set $R_c$ by assuming $R_c / (R_c + R_p) = 5.6 \%$, following the approach of \citet{Bonfanti2019Patient-specificDatasets}, and then fine-tune its value. The resulting parameters are listed in Table \ref{tab:WK2_params}.} The radius of the viscous sponge layer, $R_v$, is set to 0.03 m at all outlets so that it encloses two times the largest outlet, which has an equivalent radius of 0.0115 m.

\subsubsection{Test 1: \rev{Coupled Simulations of Left Ventricle and Cylinder}} \label{sec:test1}
In the first test, we verify the coupling scheme, focusing on its implementation and computational aspects. Specifically, we check whether the coupling occurs at the correct time steps and whether the values exchanged at the coupling interface are consistent in Alya and HemeLB \rev{and perform a convergence study}. The simulation domains include the same left ventricle geometry as in Test 0 and a cylinder representing the aorta.

A physiological cardiac output poses tight constraints on the parameters of the fluid flow simulation. \rev{First, for a given flow rate $\Qin$, the flow velocity $\bm{U}$ is inversely proportional to the inlet area. Second, from equation \ref{eq:Mach}, the Mach number is proportional to the flow velocity $\bm{U}$, the time step size $\Delta t^\mathcal{H}$, and the inverse of the voxel size $\Delta x^\mathcal{H}$. Third, from equation \ref{eq:kinematic_viscosity}, for a given kinematic viscosity $\nu_0$, the non-dimensional relaxation time $\tau_0/\Delta t^\mathcal{H}$ scales linearly with the time step size and inversely with the square of the voxel size:
\begin{equation}
    \frac{\tau_0}{\Delta t^\mathcal{H}} = 3 \nu_0 \frac{\Delta t^\mathcal{H}}{(\Delta x^\mathcal{H})^2} + \frac{1}{2}.
\end{equation}
These relations result in the competing conditions that
\begin{enumerate}[label=(\roman*)]
    \item decreasing the inlet area will increase the Mach number;
    \item decreasing the voxel size will increase the Mach number and the relaxation time; and
    \item decreasing the time step size will decrease the Mach number and the relaxation time.
\end{enumerate}
However, the Mach number must remain below 0.3 to minimise the compressibility errors \cite{Kruger2017TheMethod}. Additionally, for numerical stability, the LBGK model requires that $\tau_0/\Delta t^\mathcal{H}$ does not approach 0.5 \cite{Kruger2017TheMethod}.}

\begin{figure}[htb!]
\centering
\stackinset{l}{180pt}{t}{100pt}{\Large $\boxed{\domainheme}$}{%
\stackinset{l}{-5pt}{t}{60pt}{\Large ${\gammainheme}$}{%
\stackinset{l}{270pt}{t}{70pt}{\Large $\gammaoutheme$}{%
\stackinset{l}{85pt}{t}{5pt}{\Large $\gammawallheme$}{%
\includegraphics[width=0.6\textwidth]{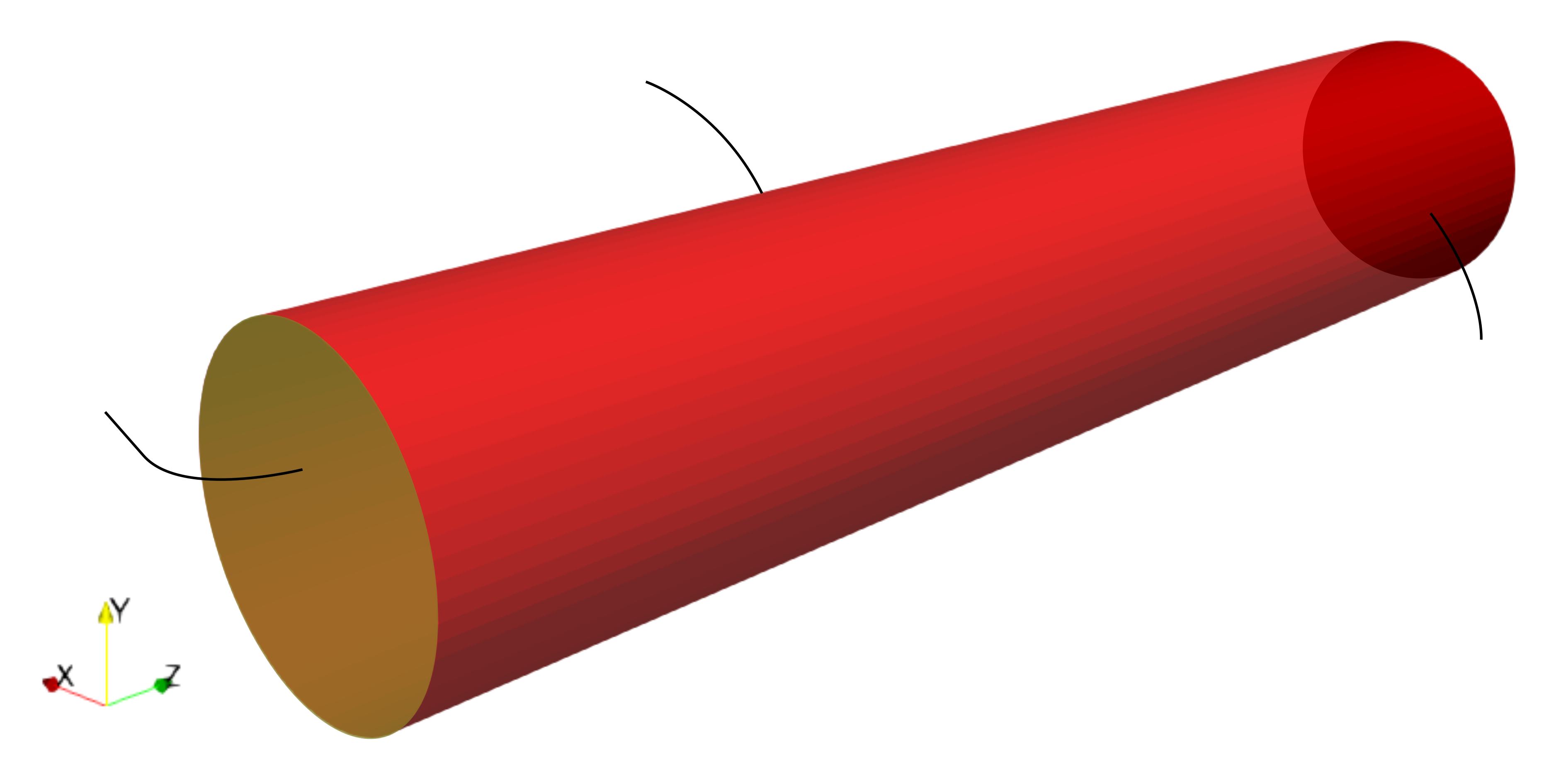}
}}}}
\caption{The cylinder model for HemeLB in Test 1. The fluid domain $\domainheme$ is the interior of this model bounded by the wall $\gammawallheme$, an inlet $\gammainheme$, and an outlet $\gammaoutheme$.}
\label{fig:cylinder}
\end{figure}

Considering the above constraints, we adopt the following choice of parameters. A cylinder of radius 0.02 m and a height of \rev{0.4} m is used, as shown in Figure \ref{fig:cylinder}. It is voxelised to a resolution ($\Delta x^\mathcal{H}$) of $1.67 \cdot 10^{-4}$ m, \rev{resulting in about 120 lattice sites along a radius and 108 million lattice sites in the entire simulation domain.} The time step size $\Delta t^\mathcal{H}$ is varied between $2.5 \cdot 10^{-5}$ s and $1.0 \cdot 10^{-5}$ s to check the implementation of the coupling scheme. These values correspond to coupling \rev{frequencies} of $k = \{4, 5, 8, 10\}$ given the time step size of $10^{-4}$ s for Alya. In particular, we study the solutions obtained from the simulation with $\Delta t^\mathcal{H} = 2.5 \cdot 10^{-5}$ s, \rev{which corresponds to $\tau_0 / \Delta t^\mathcal{H} = 0.509$}. For a physiological cardiac output resulting in a maximum flow rate $\Qin$ of around \rev{0.45} dm$^3$/s \cite{Hammermeister1974TheDisease}, the maximum velocity assuming a quadratic profile is \rev{0.716} m/s. The corresponding Mach number of \rev{0.186} is considered to contribute to a small compressibility error \cite{Kruger2017TheMethod}.

In addition, the parameters of the \rev{WK3} model at the outlet are set to \rev{$R_c = 2.0 \cdot 10^6$ kg m$^{-4}$ s$^{-1}$, $R_p = 1.0 \cdot 10^9$ kg m$^{-4}$ s$^{-1}$, and $1.5 \cdot 10^{-8}$ m$^4$ s$^2$ kg$^{-1}$}. The viscous sponge layer enclosing the outlet is set to have a radius of $R_v = 0.04$ m, which is two times that of the cylinder.

To check that the data exchanged, $\pendo$ and $\Qin$, are consistent in Alya and HemeLB, we compare their values at the coupling interface. In particular, we compare the output values of these variables in post-processing instead of the values exchanged during the coupling to ensure that the values are not only transmitted correctly but also implemented correctly as the boundary conditions in the models.

\rev{Moreover, we perform a convergence study to assess the impact of grid resolution on the solutions of the coupled model. To evaluate error convergence, we compare the pressure and flow rate at the inlet at the ejection peak (0.865 s), using the solutions on the finest grid as a reference. The absolute differences from these reference solutions are denoted as $\varepsilon_P$ for the pressure and $\varepsilon_Q$ for the flow rate.}

\rev{Grid refinement for lattice Boltzmann simulations requires a coordinated reduction of the time step size ($\Delta t^\mathcal{H}$) and lattice spacing ($\Delta x^\mathcal{H}$) to ensure numerical stability and physical consistency. Two principal scaling approaches exist \cite{Kruger2017TheMethod}: acoustic scaling maintains a constant $\Delta t^\mathcal{H} / \Delta x^\mathcal{H}$ ratio to preserve the Mach number (see equation \ref{eq:Mach}) and wave propagation characteristics, while diffusive scaling maintains a constant $\Delta t^\mathcal{H} / (\Delta x^\mathcal{H})^2$ ratio to preserve the non-dimensional relaxation time $\tau_0 / \Delta t^\mathcal{H}$ and kinematic viscosity $\nu_0 \Delta t^\mathcal{H} / (\Delta x^\mathcal{H})^2$ (see equation \ref{eq:kinematic_viscosity}). We adopt acoustic scaling because it ensures a consistent representation of pulsatile flow phenomena across refinement levels. In addition, we maintain the coupling frequency $k = 4$, which is not a grid parameter, by proportionally reducing $\Delta t^\mathcal{A}$ while keeping $\Delta x^\mathcal{A}$ fixed. The convergence properties of Alya under grid refinement have been analysed in previous standalone studies \cite{casoni2015alya, Santiago2018FullySupercomputers}, demonstrating its numerical accuracy across a range of spatial resolutions.}

\rev{Due to the substantial memory demands of voxelisation, generating the cylinder domain with an arbitrarily fine spatial resolution is impractical. To obtain more data points while managing computational constraints, we adopt a refinement ratio of 0.75 instead of the commonly used 0.5 (grid doubling). This choice remains viable, as we compare spatially averaged quantities (i.e. time-dependent only) rather than pointwise values across different resolutions.}

\subsubsection{Test 2: \rev{A Coupled Simulation of Left Ventricle and Thoracic Aorta}} \label{sec:test2}
In the second test, we evaluate the coupled model in a physiologically sound setting \rev{and assess the impact of coupling on the individual models. We use the same simulation setup as in Test 0 with the left ventricle geometry for Alya and the thoracic aorta geometry for HemeLB and couple the two codes.} The coupling frequency is set to $k = 16$ with $\Delta t^\mathcal{A} = 10^{-4}$ s and $\Delta t^\mathcal{H} = 6.25 \cdot 10^{-6}$ s.

\rev{To assess the impact of coupling on the simulated outcomes, we compare the displacement of heart muscles, $\displacement$, and the time-averaged wall shear stress (TAWSS) in the aorta between the coupled simulation and the standalone Alya and HemeLB simulations in Test 0. We compute the percentage differences in these quantities using the following formula:
\begin{equation} \label{eq:percent_diff}
    \text{\% Difference} = \bigg| \frac{\text{Current value} - \text{Reference value}}{\text{Reference value}} \bigg| \times 100 \%.
\end{equation}
TAWSS is a key biomechanical parameter that quantifies the shear forces exerted by blood flow on vessel walls. Abnormal TAWSS levels are associated with the progression of various vascular diseases, including aneurysms, atherosclerosis, and thrombosis \cite{Meng2014HighHypothesis, Staarmann2019ShearReview}. Accurately capturing TAWSS is therefore essential for characterising the haemodynamic environment and evaluating potential risks. The computation of TAWSS in HemeLB is described in our previous work \cite{Lo2024UncertaintySimulations}.}

\subsubsection{\rev{Test 3: Impact of Myocardial Scarring on Downstream Vascular Flow}}
\begin{figure}
    \centering
    \includegraphics[width=\linewidth]{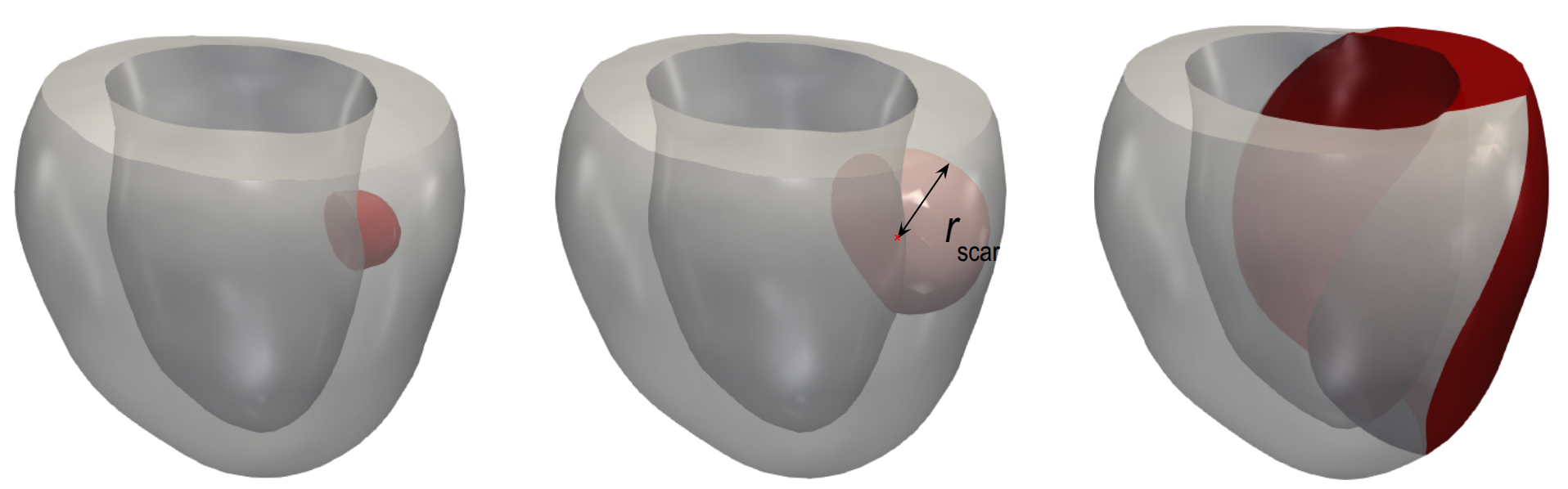}
    \caption{\rev{Left ventricle with idealised scars of different sizes: $r_\mathrm{scar} = $ 0.5 cm, 1.0 cm, and 2.5 cm (left, middle, and right panels, respectively).}}
    \label{fig:scars-methodology}
\end{figure}

\rev{As a final test, we evaluate whether the model can capture physiologically relevant features of disease. Specifically, we investigate the impact of myocardial scarring in the left ventricle on aortic haemodynamics by repeating the coupled simulation from Test 2 under pathological conditions. To this end, we generate idealised scars of varying sizes, as illustrated in Figure \ref{fig:scars-methodology}. Scars are created by selecting a point on the endocardial surface, computing the Euclidean distance from this point to each mesh node, and applying a threshold ${r}_\mathrm{scar}$ of 0.5 cm, 1.0 cm, and 2.5 cm to define the scar region. The scarred regions were considered electrically non-conductive: the cellular model is deactivated ($\Iion = 0$ in equation \ref{eq:mono-parabolic-diff}), and the monodomain diffusivity $\bm D$ is set to zero \cite{arevalo2016arrhythmia, salvador2022role}. By varying the size of the scar, we aim to assess how different degrees of cardiac dysfunction influence haemodynamic outcomes. In particular, we compare the TAWSS distributions in the aorta with the distribution of Test 2 using the percentage difference defined in equation \ref{eq:percent_diff}.}

%% file: 3_Results-Discussion.tex
\section{Results and Discussion} \label{sec:results_discussion}
In this section, we present and analyse the results of the \rev{four} simulation tests in Sections \ref{sec:results_test0} -- \ref{sec:results_test3}. Lastly, we assess the computational efficiency of the coupled model. All our simulations are performed on the ARCHER2 UK national supercomputer (\url{https://www.archer2.ac.uk}).

\subsection{Test 0: Baseline Solutions for Left Ventricle and Thoracic Aorta} \label{sec:results_test0}
\begin{figure}[!htb]
    \centering
    \begin{subfigure}{0.48\linewidth}
        \centering
        \includegraphics[width=\linewidth]{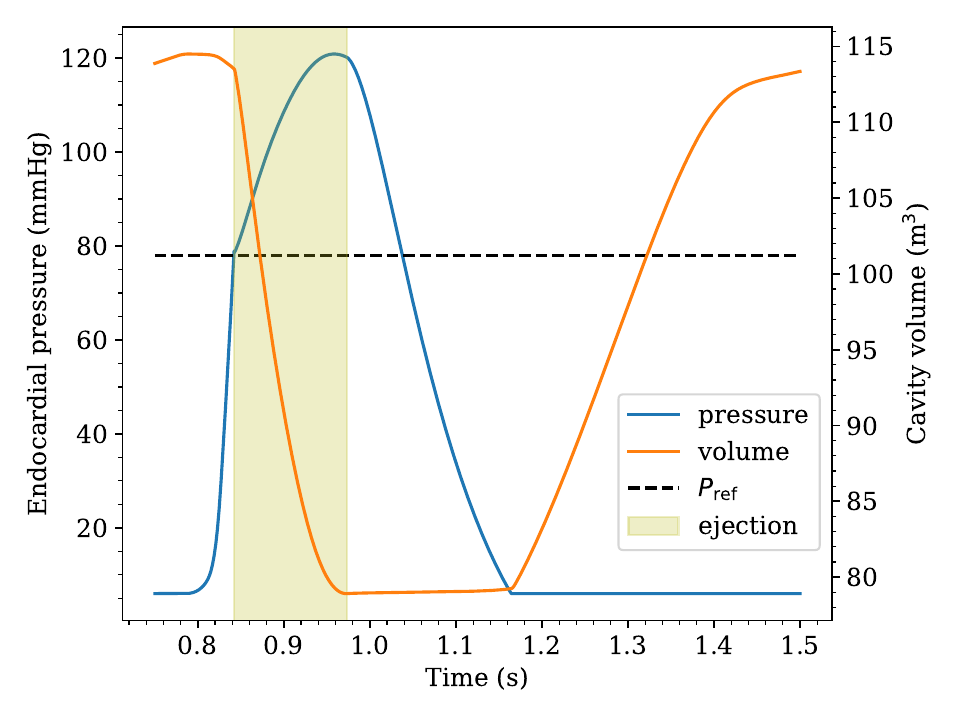}
        \caption{Left ventricular endocardial pressure and volume}
    \end{subfigure}
    \hfill
    \begin{subfigure}{0.48\linewidth}
        \centering
        \includegraphics[width=\linewidth]{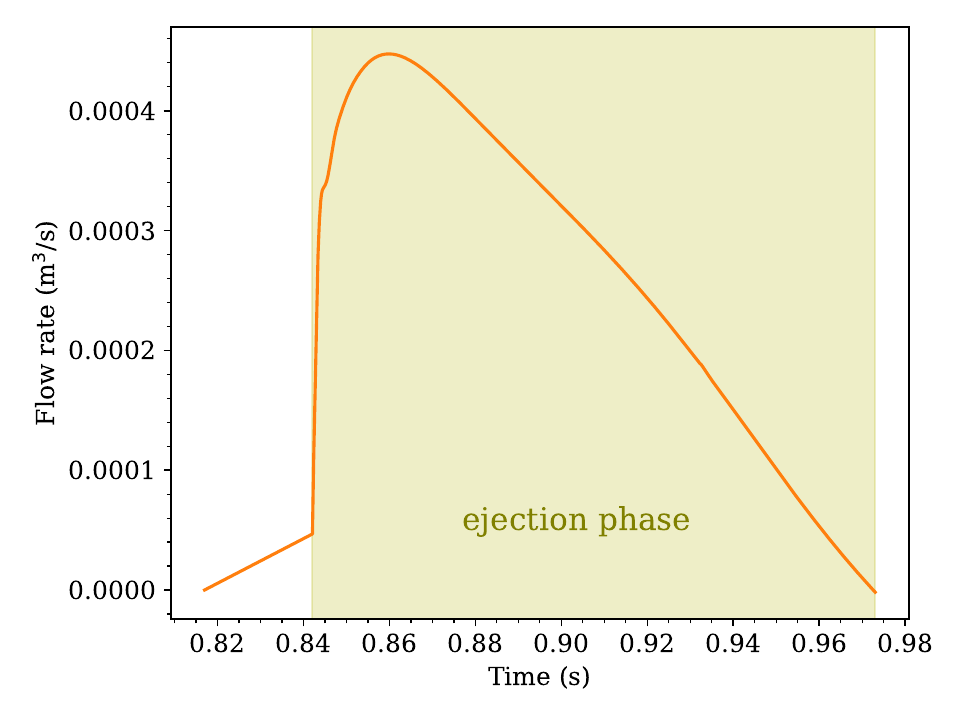}
        \caption{Aortic inflow rate}
    \end{subfigure}
    \caption{Results of Test 0. (a) Left ventricular endocardial pressure and volume from the standalone Alya simulation. The endocardial pressure immediately before the ejection phase suggests a reference pressure $P_\text{ref}$ of \rev{78} mmHg for the fluid domain in the subsequent tests. \rev{(b) Inlet flow rate profile prescribed in the standalone HemeLB simulation, obtained as the negative derivative of the ventricular volume time series during the ejection phase with a prepended warm-up period (see equation \ref{eq:coupled-kinematic}).}}
    \label{fig:test0_alya-hemelb}
\end{figure}

We run the standalone simulation of Alya for two heartbeats and present results from the second heartbeat, as the electromechanical model typically requires a few beats to reach a periodic solution. \rev{Figure \ref{fig:test0_alya-hemelb}a shows the time series of the left ventricular endocardial pressure and volume, with the ejection phase occurring from $t = 0.842$ s to $t = 0.973$ s.} The endocardial pressure immediately before the ejection phase suggests a reference pressure $P_\text{ref}$ of \rev{78} mmHg for the fluid domain in the subsequent tests.

\rev{The ventricular volume in the ejection phase is used to derive the inlet flow rate profile for the fluid domain, as shown in Figure \ref{fig:test0_alya-hemelb}b, following equation \ref{eq:coupled-kinematic}. To ensure a smooth transition, a warm-up period of 0.025 s is prepended to the inflow profile, during which the flow rate increases steadily to its initial value. We run the standalone simulation of HemeLB with this inlet profile. Figures \ref{fig:test0-2_displ}a and \ref{fig:test0-2_TAWSS}a show the magnitude of heart muscle displacement and the time-averaged wall shear stress in the aorta, respectively, for comparison with Test 2.}

\subsection{Test 1: \rev{Coupled Simulations of Left Ventricle and Cylinder}} \label{sec:results_test1}
Here we present the results of the coupled simulations using the left ventricle and cylinder geometries. First of all, we assess the reliability of the coupling scheme. We perform the coupled simulations using different time step sizes and coupling frequencies in HemeLB while keeping them fixed in Alya. These simulations are completed at the specified simulation times, showing that our implementation of the coupling algorithm is reliable.

\begin{figure}[!htb]
    \centering
    \includegraphics[width=0.6\linewidth]{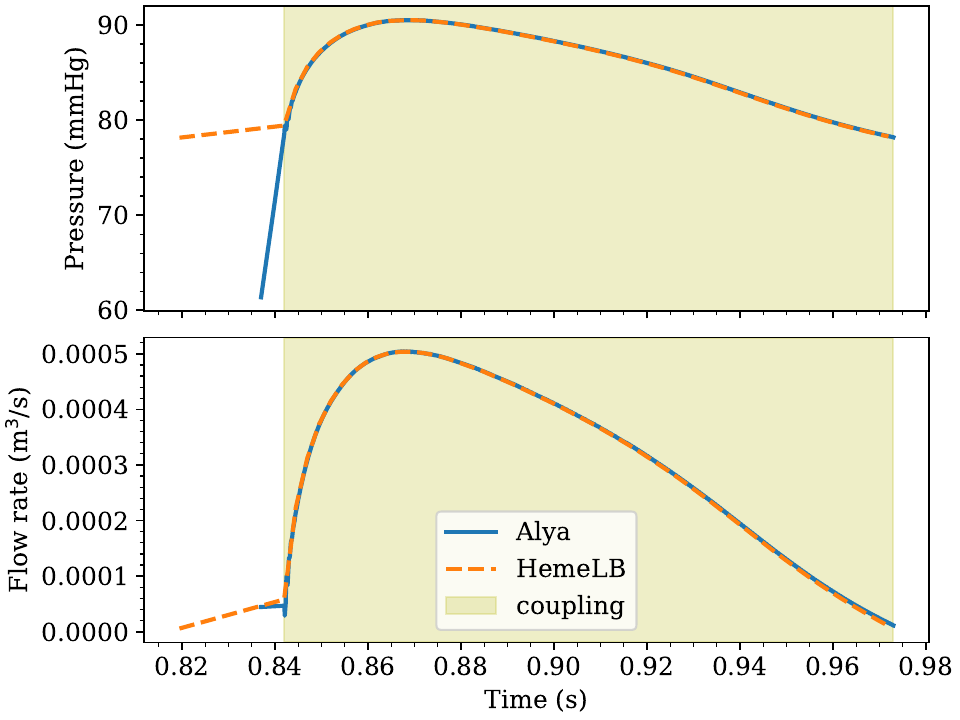}
    \caption{Pressure and flow rate values at the coupling interface \rev{from Alya and HemeLB, demonstrating consistency between} the two codes during coupling. \rev{The results are obtained from the coupled simulation with $k = 4$ in Test 1.}}
    \label{fig:test1_alya-hemelb}
\end{figure}

Second, we check the consistency of the exchanged data. \rev{During the coupling phase, the pressure and flow rate values at the coupling interface from Alya and HemeLB show excellent agreement. As an example, the outputs for $k=4$ are presented in Figure \ref{fig:test1_alya-hemelb}. This suggests that the exchanged data are consistent in the two solvers and that our coupling algorithm is correctly implemented.}

\begin{figure}[!htb]
    \centering
    \begin{subfigure}{0.48\linewidth}
        \centering
        \includegraphics[width=\linewidth]{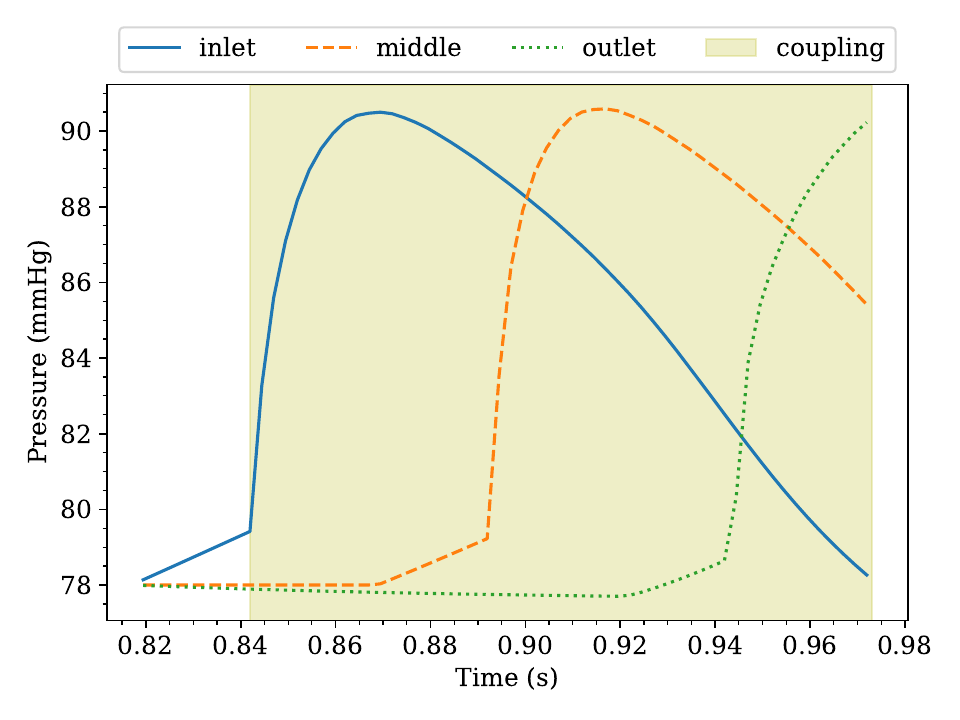}
        \caption{Pressure}
    \end{subfigure}
    \hfill
    \begin{subfigure}{0.48\linewidth}
        \centering
        \includegraphics[width=\linewidth]{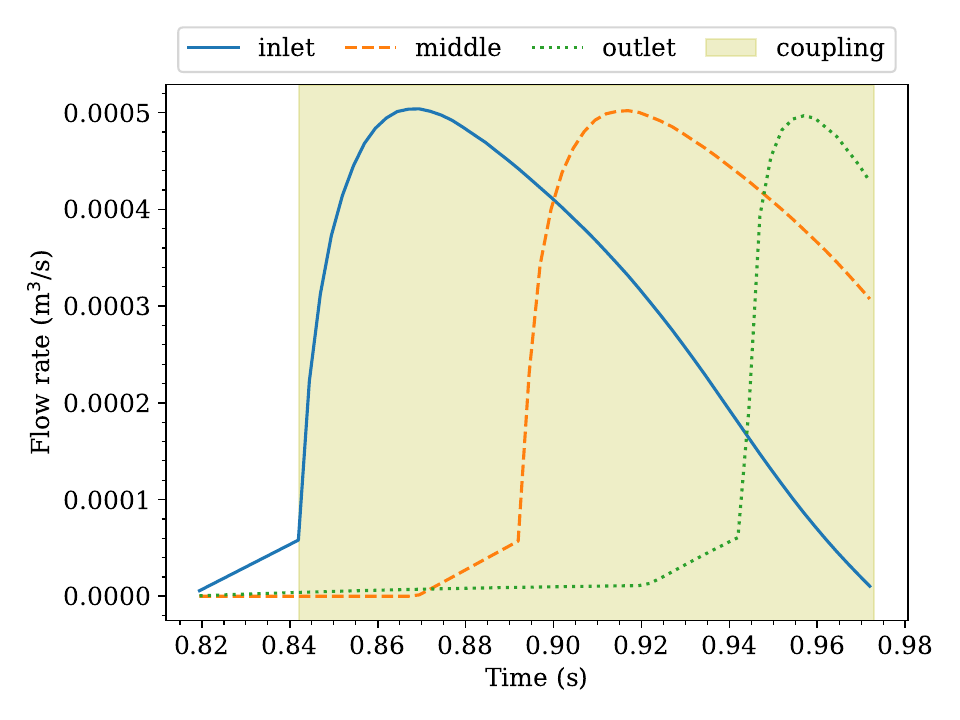}
        \caption{Flow rate}
    \end{subfigure}
    \caption{(a) Pressure and (b) flow rate on the inlet, middle, and outlet planes of the cylinder domain \rev{in Test 1. These variables remain bounded and free of spurious oscillations, demonstrating the stability of the coupled framework.}}
    \label{fig:test1_hemelb_1D}
\end{figure}

\rev{Third, we evaluate the stability of the coupled simulations. Instabilities are more likely to arise in the fluid dynamics simulations because the flow variables are highly sensitive to numerical oscillations. Therefore, we focus on assessing stability through the pressure and flow rate in the HemeLB simulations. We find that these variables remain bounded and free of spurious oscillations at the inlet, middle, and outlet of the fluid domain throughout the simulations. As an example, Figure \ref{fig:test1_hemelb_1D} presents the results for $k = 4$. The same is observed for the electromechanical solutions of Alya. These results confirm the stability of the coupled framework.}

\begin{figure}[!htb]
    \centering
    \begin{subfigure}{0.48\linewidth}
        \centering
        \includegraphics[width=\linewidth]{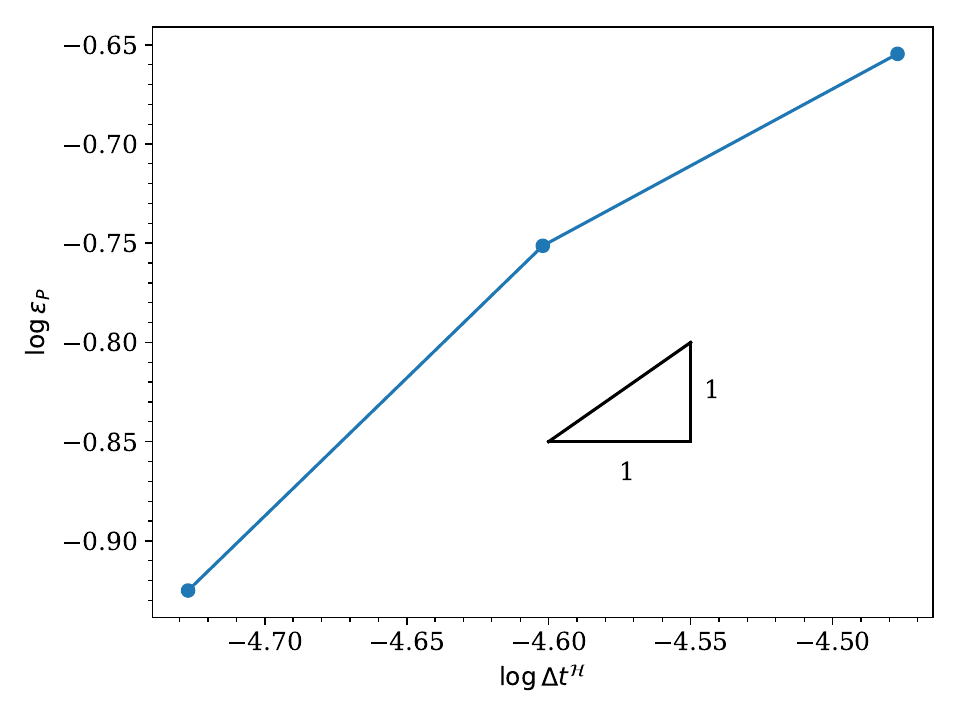}
        \caption{Pressure}
    \end{subfigure}
    \hfill
    \begin{subfigure}{0.48\linewidth}
        \centering
        \includegraphics[width=\linewidth]{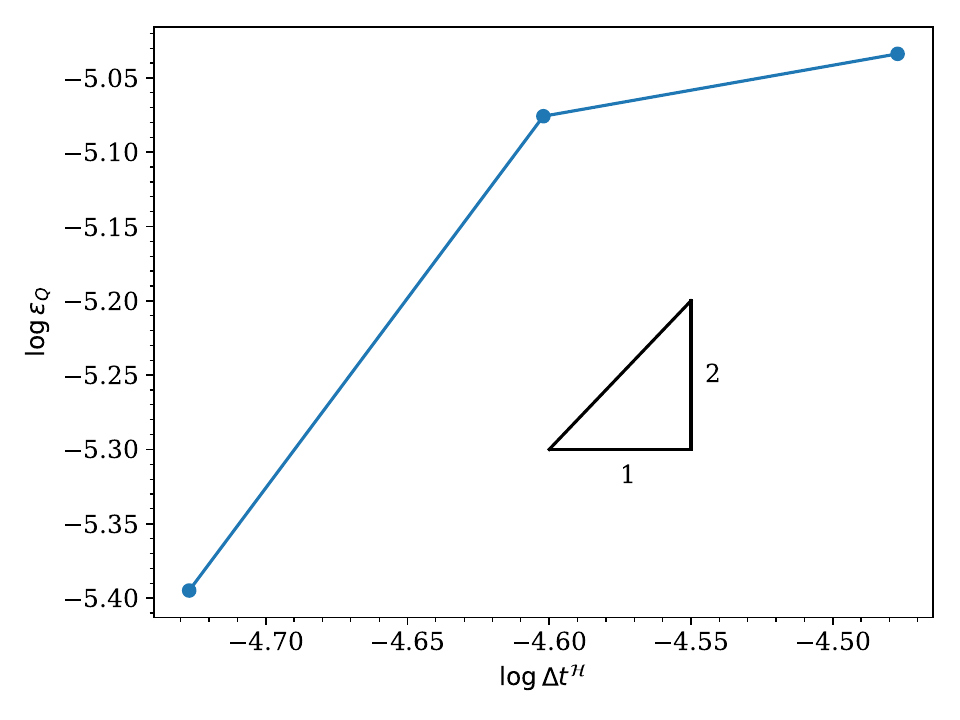}
        \caption{Flow rate}
    \end{subfigure}
    \caption{\rev{Convergence of errors in (a) pressure ($\varepsilon_P$) and (b) flow rate ($\varepsilon_Q$) in the coupled Alya-HemeLB simulation of Test 1. The pressure and flow rate are measured at the inlet at the ejection peak (0.865 s) across four resolutions. Their absolute errors are computed relative to the solutions on the finest grid ($\Delta x^\mathcal{H} = 9.375 \cdot 10^{-5}$ m and $\Delta t^\mathcal{H} = 1.40625 \cdot 10^{-5}$ s). The observed convergence rates are approximately one for the pressure and two for the flow rate.}}
    \label{fig:test1_convergence}
\end{figure}

\rev{Fourth, we assess the impact of grid resolution on the solutions of the coupled model. The absolute errors in pressure and flow rate are shown in Figure \ref{fig:test1_convergence}. The observed convergence rates are approximately one for the pressure and two for the flow rate. We note that the differences in discretisation schemes between the lattice Boltzmann and finite element methods may affect error propagation, potentially influencing the observed convergence rates. Despite this, the second-order convergence of the flow rate suggests that the overall flow dynamics remain well captured across different grid resolutions.}

\subsection{Test 2: \rev{A Coupled Simulation of Left Ventricle and Thoracic Aorta}} \label{sec:results_test2}
Here we present the results of the \rev{coupled} simulations using the left ventricle and thoracic aorta geometries. Initially, the coupled simulations become unstable at the tested coupling frequencies ($k = 8, 10, 12, 14, 16, 20$). However, applying the low-pass filter with an appropriate smoothing factor ($\alpha$), all these simulations become stable. This demonstrates the effectiveness of the low-pass filter in stabilising the coupled model for this specific system. In the following, we look into the simulation using $k = 16$ and $\alpha = 0.05$. The maximum Mach number in this simulation is \rev{0.148}, remaining below the recommended limit to ensure acceptable compressibility errors \cite{Kruger2017TheMethod}.

\begin{figure}[!htb]
    \centering
    \begin{subfigure}{0.48\linewidth}
        \centering
        \includegraphics[width=\linewidth]{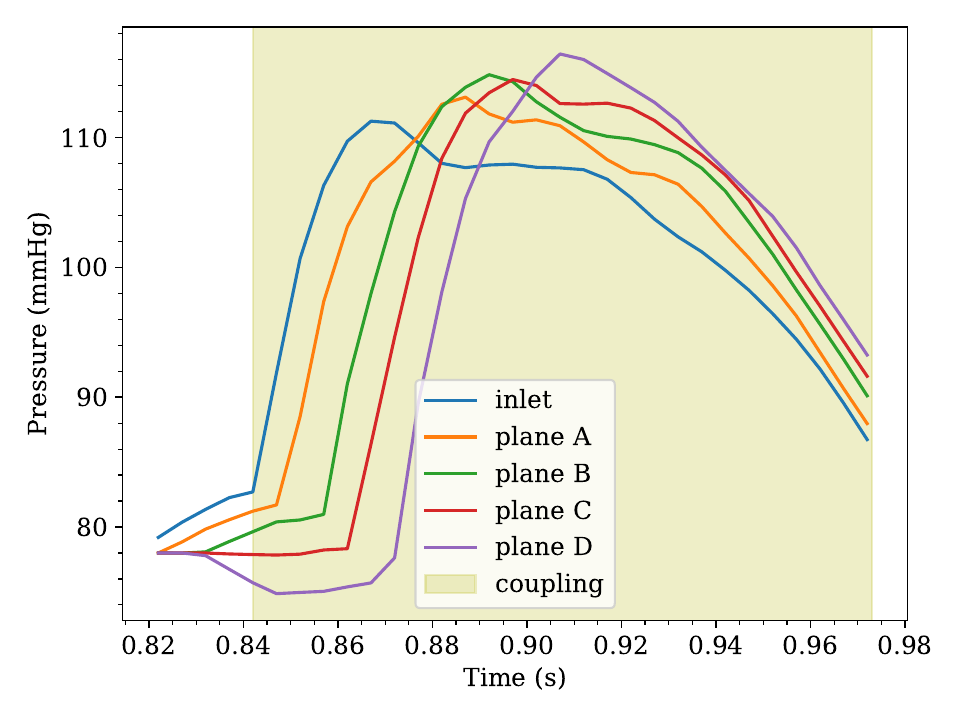}
        \caption{Pressure}
    \end{subfigure}
    \hfill
    \begin{subfigure}{0.48\linewidth}
        \centering
        \includegraphics[width=\linewidth]{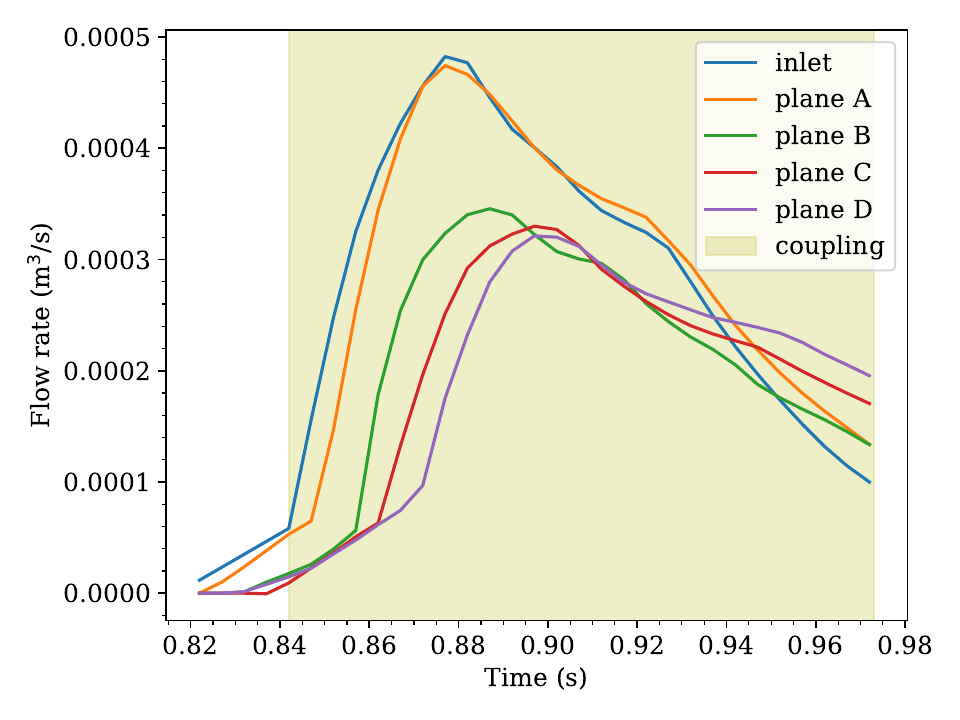}
        \caption{Flow rate}
    \end{subfigure}
    \caption{\rev{(a) Pressure and (b) flow rate at the inlet and planes A to D (see Figure \ref{fig:test2_velocityOnPlanes} for their locations) of the aorta in Test 2.}}
    \label{fig:test2_hemelb_1D}
\end{figure}

\begin{figure}[!htb]
    \centering
    \includegraphics[width=\linewidth]{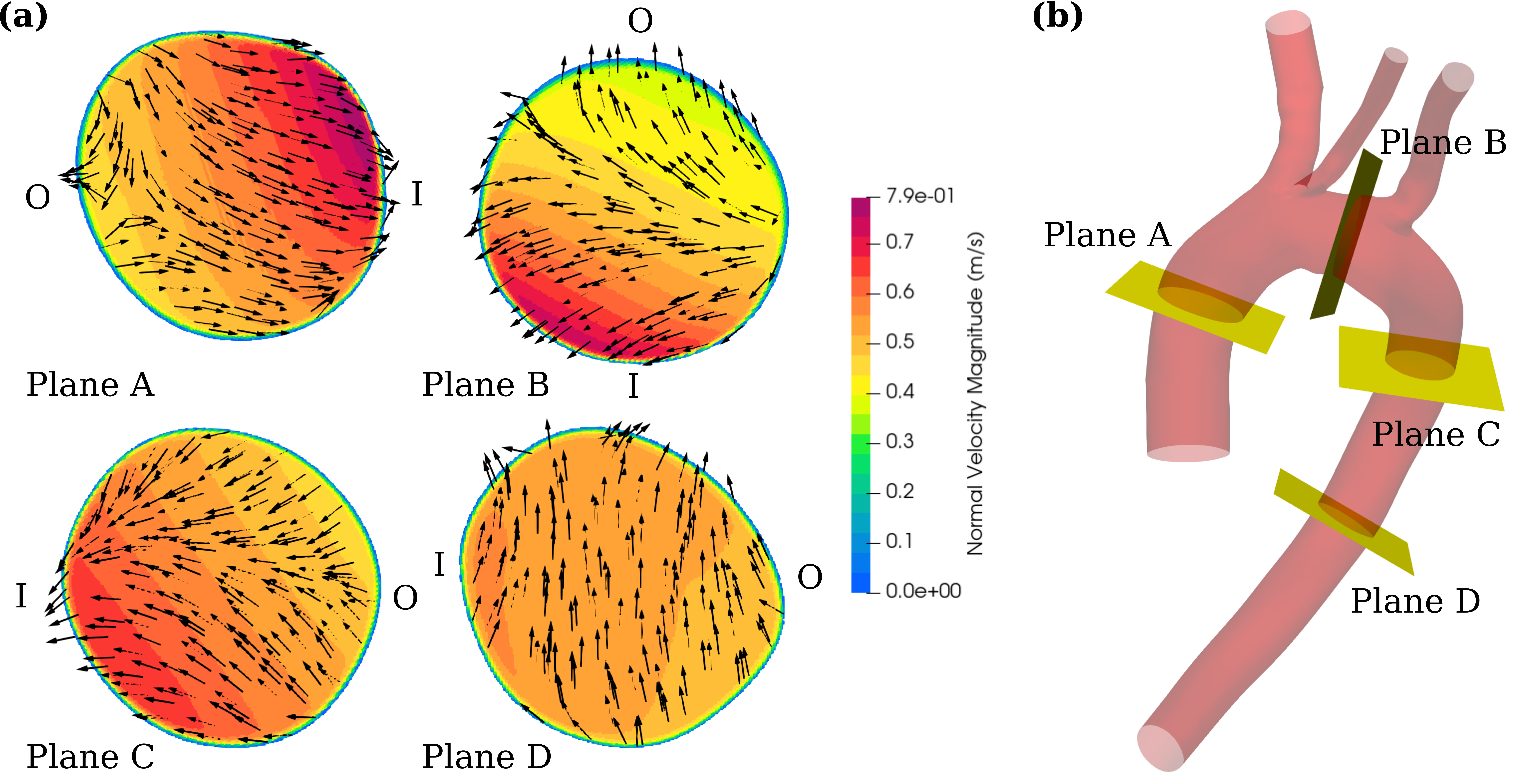}
    \caption{(a) Flow velocity at the ejection peak ($t = 0.88$ s) on (b) different cross-sections of the aorta \rev{in Test 2}. The heat map illustrates the magnitude of the normal velocity, while the arrows indicate the directions of the tangential velocity. The labels ``O" and ``I" indicate the outer and inner sides of the aorta, respectively.}
    \label{fig:test2_velocityOnPlanes}
\end{figure}

\rev{To characterise the flow behaviour, we analyse the pressure and flow rate at the aortic inlet and cross-sectional planes A to D (see Figure \ref{fig:test2_velocityOnPlanes} for their locations). Figure \ref{fig:test2_hemelb_1D} presents the acquired time series. Both variables show amplitudes and waveforms consistent with \textit{in vivo} measurements \cite{2009CardiovascularSystem, Vennin2015NoninvasiveConcept, Qasem2008DeterminationPulse, Alastruey2016OnHaemodynamics}. The successive peaks from the inlet to plane D reflect wave propagation effects similar to those observed in human physiology \cite{Mattace-Raso2010DeterminantsValues}.} Wave propagation is simulated because the LBM solves for a weakly compressible flow, allowing for slight changes in fluid density (see equation \ref{eq:lb-density}).

\rev{For a more detailed understanding of the flow dynamics, we visualise the velocity field on planes A to D. Figure \ref{fig:test2_velocityOnPlanes} shows the magnitude of the normal velocity and the directions of the tangential velocity at the ejection peak ($t = 0.88$ s). Across all planes, the flow directions align with the vessel morphology. On planes A and C, the flow accelerates along the inner bend, evidenced by the higher normal velocities. Plane B exhibits flow splitting, with one component directed upward towards the neck vessels and the other downward into the descending aorta. These results are comparable to those observed at the peak systole in studies examining aortic blood flow \cite{Madhavan2018TheFlow, Youssefi2018ImpactAorta}.}

\begin{figure}[!htb]
    \centering
    \includegraphics[width=0.5\linewidth]{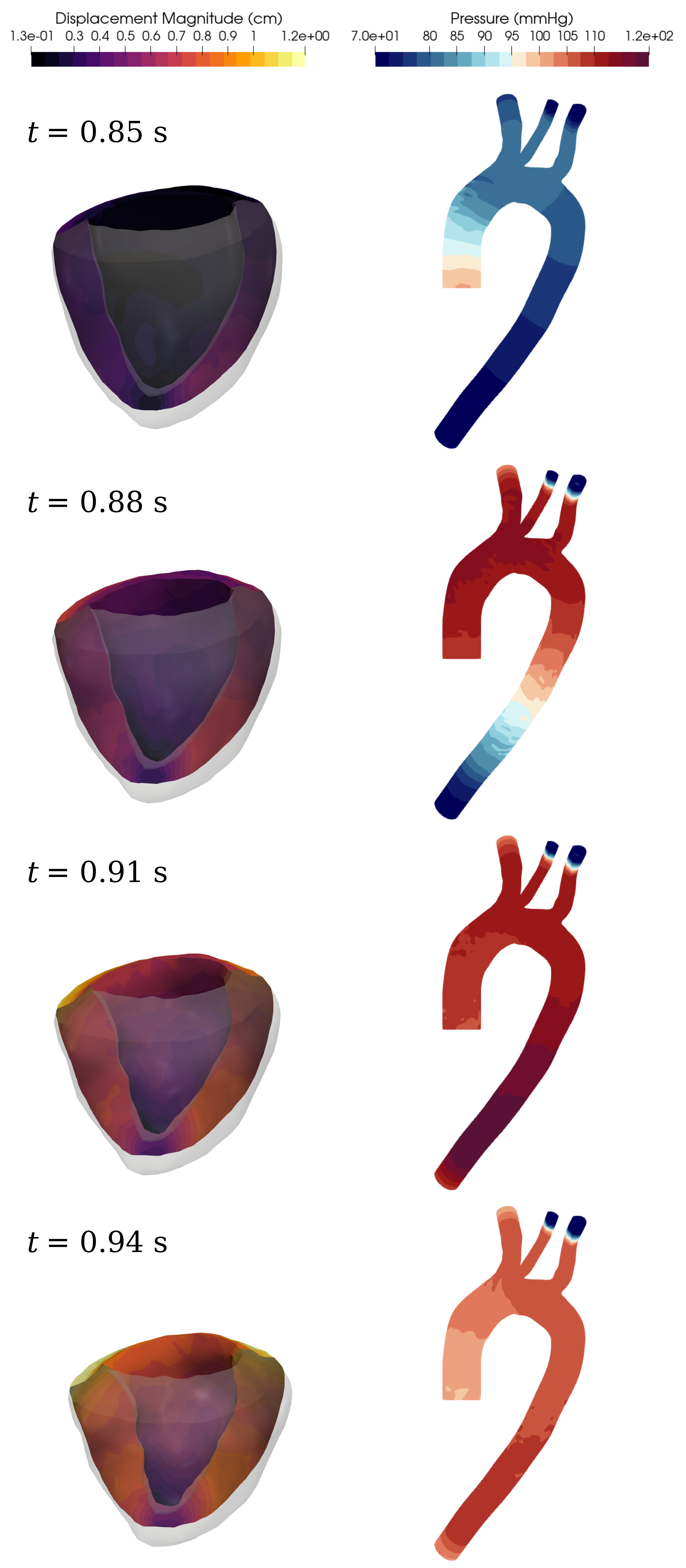}
    \caption{\rev{Displacement of heart muscles and blood pressure within the aorta during the coupling phase in Test 2. The coupled model effectively captures the 3D heart contraction and vascular blood flow as well as their dynamical interactions.}}
    \label{fig:test2_ventricle-aorta}
\end{figure}

\rev{To explore how the heart muscles and vascular blood flow interact, we visualise the displacement of heart muscles and blood pressure within the aorta during the coupling phase. As shown in Figure \ref{fig:test2_ventricle-aorta}, the muscle contraction is more pronounced at the ventricle base and decreases towards the apex.} Simultaneously, a high-pressure front originates from the inlet and propagates downstream. Around the outlets, the effects of the sponge layer are observed, where the pressure is lower than the surroundings. The coupled model effectively captures the 3D heart contraction and vascular blood flow within the vessels as well as their dynamical interactions.

\begin{figure}[!htb]
    \centering
    \includegraphics[width=\linewidth]{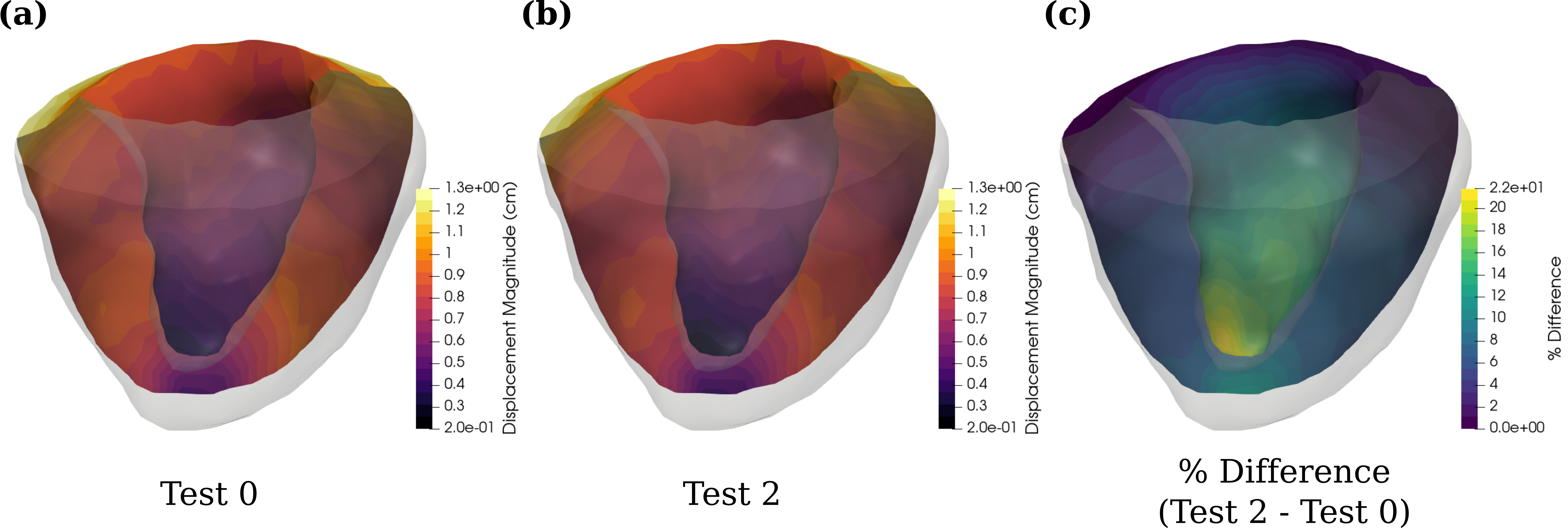}
    \caption{\rev{Magnitude of heart muscle displacement at the end of the ejection phase ($t = 0.97$ s) for (a) the standalone Alya simulation in Test 0 and (b) the coupled Alya-HemeLB simulation in Test 2, and (c) the percentage difference of Test 2 relative to Test 0.}}
    \label{fig:test0-2_displ}
\end{figure}

\rev{To assess the impact of coupling on cardiac mechanics, we compare the displacement of the heart muscles at the end of the ejection phase between Test 0 and Test 2. Figure \ref{fig:test0-2_displ}a shows the displacement from the standalone Alya simulation (Test 0), while Figure \ref{fig:test0-2_displ}b presents the results from the coupled simulation (Test 2). Both exhibit a similar distribution of muscle displacement; however, there are notable differences in magnitude. As shown in Figure \ref{fig:test0-2_displ}c, the percentage difference between the two solutions increases from the base to the apex and from the epicardium to the endocardium, peaking at 22 \% close to the apex at the endocardium. These differences underscore the importance of accurately prescribing endocardial pressure, which is determined by vascular blood flow.}

\begin{figure}[!htb]
    \centering
    \includegraphics[width=\linewidth]{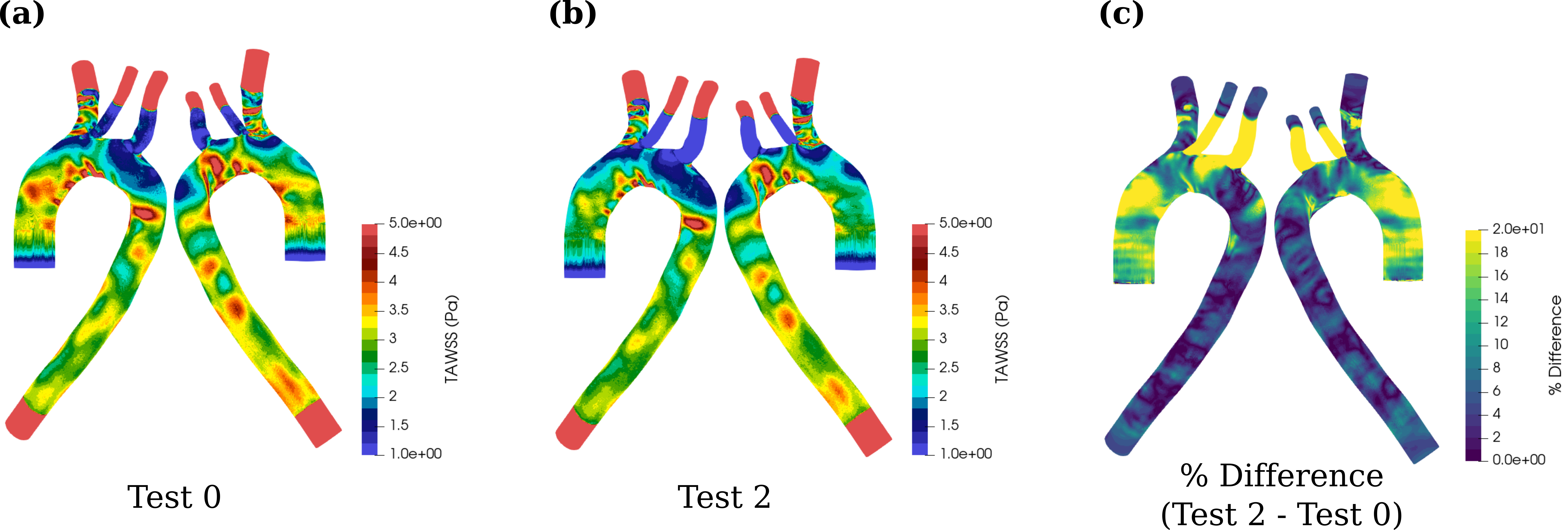}
    \caption{\rev{Time-averaged wall shear stress (TAWSS) in the aorta during the ejection phase for (a) the standalone HemeLB simulation in Test 0, (b) the coupled Alya-HemeLB simulation in Test 2, and (c) the percentage difference of Test 2 relative to Test 0.}}
    \label{fig:test0-2_TAWSS}
\end{figure}

\rev{Similarly, we evaluate the influence of coupling on the TAWSS in the aorta during the ejection phase between Test 0 and Test 2, as shown in Figure \ref{fig:test0-2_TAWSS}. While the overall distribution of TAWSS in the standalone HemeLB simulation (Test 0) is similar to that in the coupled simulation (Test 2), notable differences in magnitude are observed. As shown in Figure \ref{fig:test0-2_TAWSS}c, the percentage difference is more pronounced in the upstream region, reaching up to 20 \% in parts of the ascending aorta. These differences highlight the critical role of accurately prescribing the inlet profile, which is governed by the cardiac output resulting from the coupling with the ventricular electromechanics.}

% Computational performance
Since Alya and HemeLB are coupled in a staggered manner, the scalability of the coupled model is entirely determined by the scalability of the individual solvers, which has been extensively demonstrated in previous studies \cite{Vazquez2016Alya:Exascale, Zacharoudiou2023DevelopmentSimulation}. The coupling scheme, which performs data exchange by reading and writing two small files (each containing only two numbers), runs sequentially. As a result, the time spent on coupling remains relatively constant and minimal. When Alya runs on two compute nodes and HemeLB on 64 nodes, the percentage of the time spent on coupling relative to the elapsed time per \rev{coupling iteration} is about 4 \% for Alya and 6 \% for HemeLB (see \ref{sec:computational_performance} for more details).

\subsection{\rev{Test 3: Impact of Myocardial Scarring on Downstream Vascular Flow}} \label{sec:results_test3}
\rev{Here we present the results of the coupled simulations with cardiac scarring and analyse the impact of the scarring on downstream vascular flow.}

\begin{figure}[!htb]
    \centering
    \includegraphics[width=\linewidth]{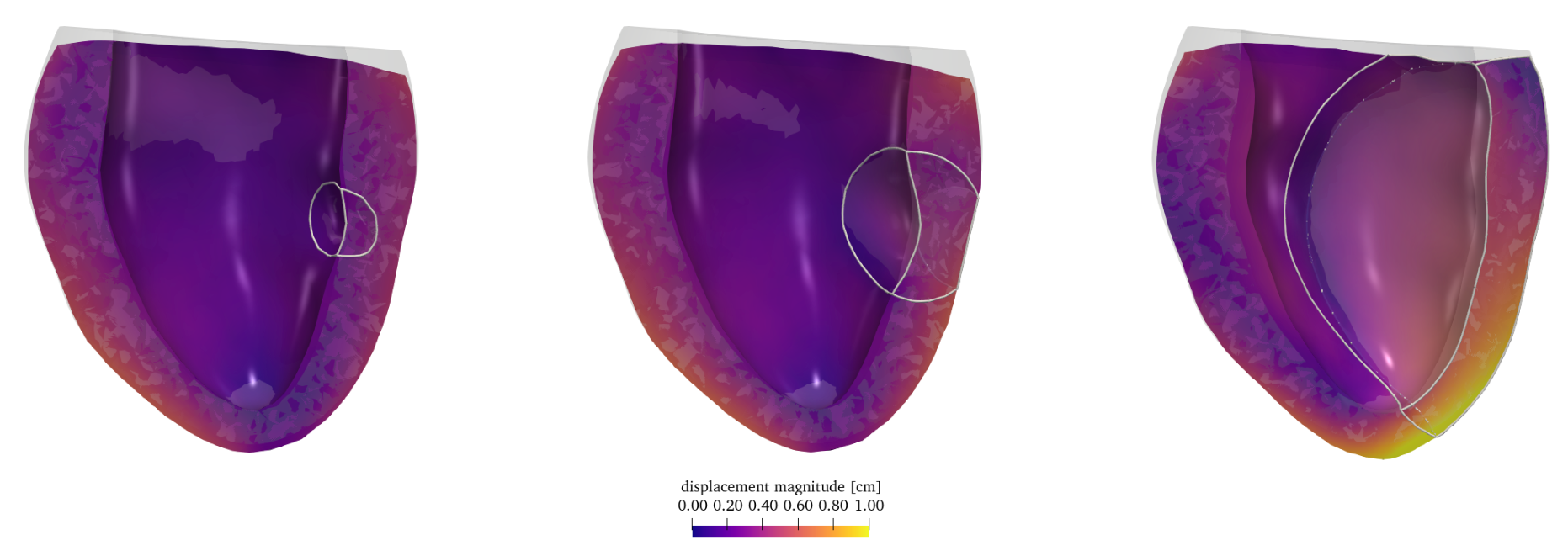}
    \caption{\rev{Left ventricular contraction at the ejection peak in Test 3, with myocardial scars of varying sizes (bounded by grey lines): 0.5 cm (left, peak at 0.88 s), 1.0 cm (middle, 0.88 s), and 2.5 cm (right, 0.93 s). The transparent overlay shows the left ventricular anatomy in its initial configuration.}}
    \label{fig:test3_scar}
\end{figure}

\begin{figure}[!htb]
    \centering
    \begin{subfigure}{0.48\linewidth}
        \centering
        \includegraphics[width=\linewidth]{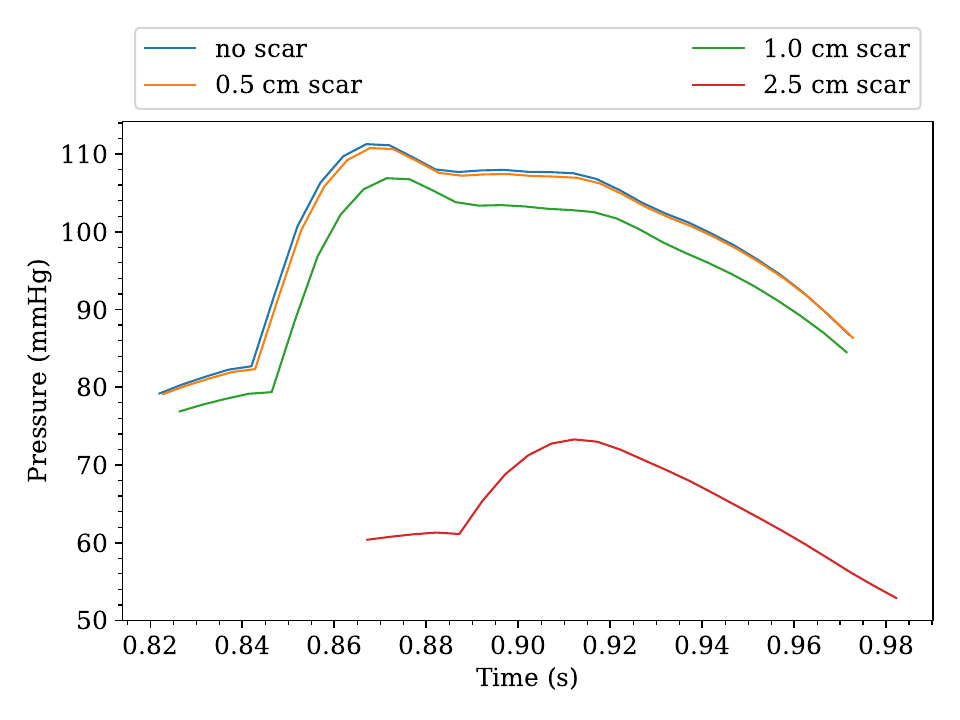}
        \caption{Pressure}
    \end{subfigure}
    \hfill
    \begin{subfigure}{0.48\linewidth}
        \centering
        \includegraphics[width=\linewidth]{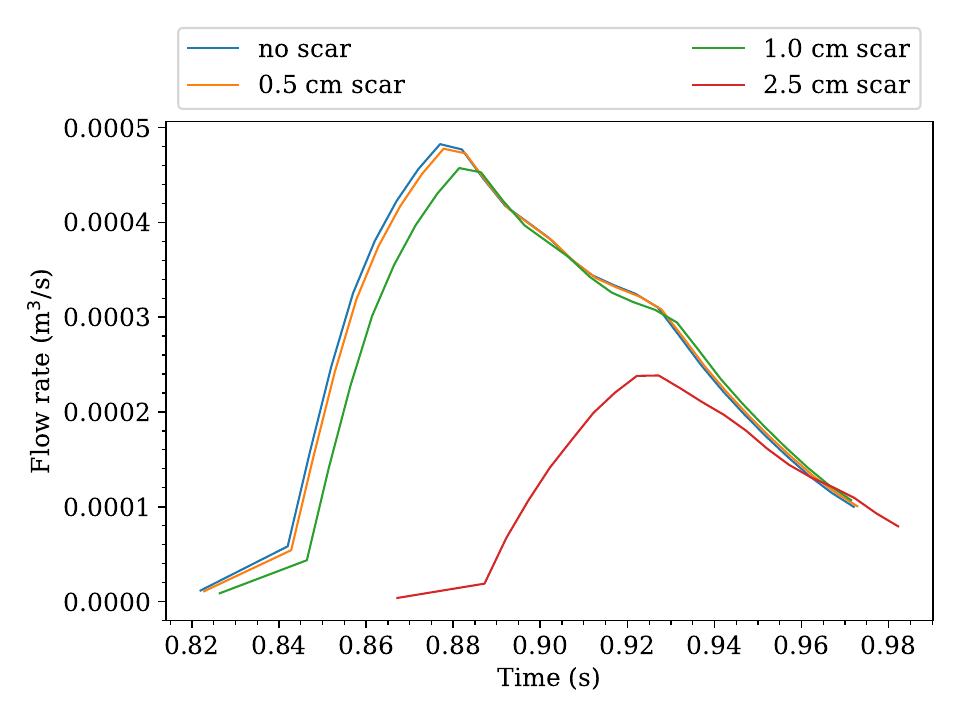}
        \caption{Flow rate}
    \end{subfigure}
    \caption{\rev{(a) Pressure and (b) flow rate at the coupling interface in the Alya-HemeLB simulations in Test 3, each using the same heart model but with scars of varying sizes, coupled with the thoracic aorta model. Compared to the scar-free case (Test 2), both the pressure and the flow rate decrease more significantly as the scar size increases, and the ejection phase starts later.}}
    \label{fig:test3_alya-hemelb}
\end{figure}

\rev{Figure \ref{fig:test3_scar} shows the contraction of the left ventricle with myocardial scars of different sizes at the ejection peak. As the scar size increases, noticeable changes in the contraction pattern emerge. To better quantify these differences, Figure \ref{fig:test3_alya-hemelb} presents the time series of the pressure and flow rate at the coupling interface for different scar sizes, alongside the scar-free case from Test 2 for comparison. As the scar size increases, both pressure and flow decrease, and the onset of the ejection phase is delayed. These deviations from the scar-free case become more pronounced with larger scars, highlighting the influence of myocardial damage on cardiac function and haemodynamics.}

\begin{figure}[!htb]
    \centering
    \includegraphics[width=\linewidth]{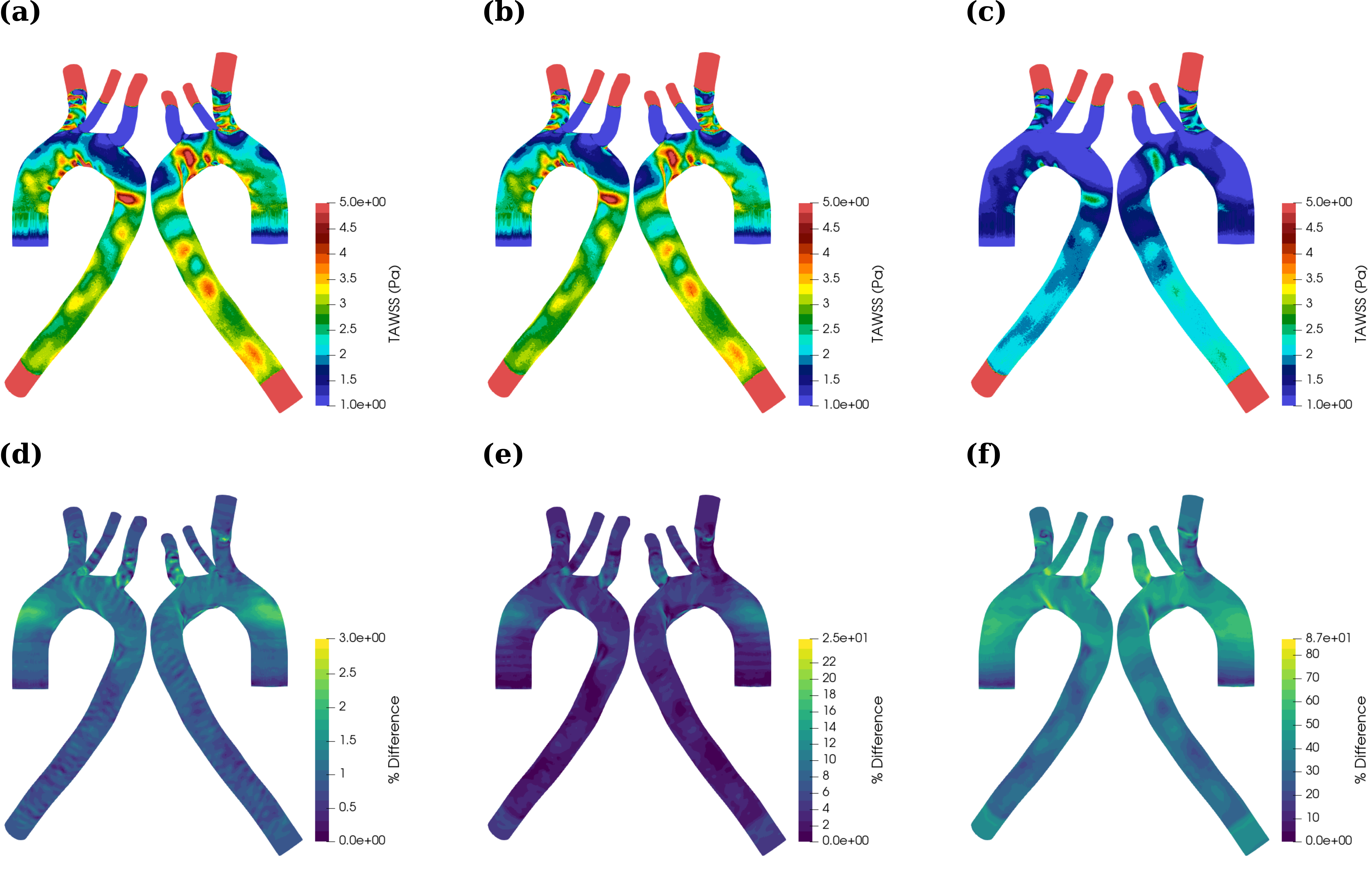}
    \caption{\rev{Time-averaged wall shear stress (TAWSS) in the aorta during the ejection phase of the coupled simulations in Test 3, using heart models with myocardial scars of different sizes: (a) 0.5 cm, (b) 1.0 cm, (c) 2.5 cm. The percentage difference of (a), (b), and (c) relative to Test 2 in Figure \ref{fig:test0-2_TAWSS}b is shown in (d), (e), and (f).}}
    \label{fig:test3_TAWSS}
\end{figure}

\rev{To further investigate the haemodynamic impact of myocardial scarring, we examine the distribution of TAWSS in the aorta during the coupling phase. Figure \ref{fig:test3_TAWSS} shows the TAWSS distributions for different scar sizes, along with the percentage difference relative to the scar-free case from Test 2 (Figure \ref{fig:test0-2_TAWSS}b). As the scar size increases, deviations in the TAWSS distribution from the scar-free case become more noticeable. The maximum difference in the aorta increases from 3.0 \% in the 0.5 cm scar case to 25 \% in the 1.0 cm scar case and reaches up to 87 \% in the 2.5 cm scar case. These results highlight the substantial impact of myocardial scarring on aortic haemodynamics, which may have important implications for vascular disease development, as abnormal TAWSS is linked to conditions such as aneurysms, atherosclerosis, and thrombosis \cite{Meng2014HighHypothesis, Staarmann2019ShearReview}.}

%% file: 4_Limitations.tex
\section{Limitations} \label{sec:limitations}
A primary limitation of the coupled model presented in this study is the inherent necessity of assuming a specific spatial velocity distribution for inflow into the vasculature (see equation \ref{eq:inlet_velocity}). This assumption is required because the flow rate provided by Alya only specifies the total normal velocity at the inlet, leaving the local direction and magnitude of the velocity unspecified. Although the parabolic-shaped profile used in this study (equation \ref{eq:parabolic_profile}) may have limited effects on the flow downstream in a healthy aorta due to the accumulated influence of vessel walls, it can significantly impact the flow near the inlet of a diseased aorta, where velocity patterns are highly complex \cite{Madhavan2018TheFlow, Youssefi2018ImpactAorta}. Ultimately, the spatial velocity profile at the aortic root should be determined by the blood flow dynamics within the heart. Achieving this goal will require incorporating a fluid dynamics model for the heart compartments, coupling the movement of the endocardium with blood flow, and mapping the flow variables between the heart compartments and vascular models. This comprehensive model can be developed through the integration of existing models using multi-component coupling.

Another limitation is the application of a low-pass filter for the flow rate (equation \ref{eq:low-pass_filter}) to stabilise the coupled simulations. This numerical technique is necessitated by two main factors. The first is the competing conditions discussed in Section \ref{sec:coupled-model}, which stem from the limitations of the LBGK model when used to simulate flows with a high Reynolds number \cite{Kruger2017TheMethod}. The second factor is the stability requirements of the coupling scheme. While a partitioned coupling scheme reduces computational complexity and cost by dividing the system into smaller subsystems and making certain assumptions, it potentially compromises the stability of the coupling \cite{Piperno1997Explicit/implicitSimulations, Kuttler2006ADomains, Bazilevs2009Patient-specificDevice}. Future studies could improve the stability by using an implicit coupling scheme and sub-iterating two biophysical models within a time step until the solutions converge. This approach may avoid the need for numerical stabilisation methods and reduce the number of assumptions required in the present work.

A third limitation is that the cardiac electromechanics and vascular haemodynamics are coupled only during the ejection phase. This is the phase during which the aortic valve is open and the flow velocity at the inlet is primarily unidirectional. In other cardiac phases, the closure of the aortic valve means that the vascular flow is governed by the residual impulse, resulting in secondary flow patterns that necessitate a more detailed analysis of the inlet velocity distribution \cite{Kilner1993HelicalMapping, EscobarKvitting2004FlowSurgery, Youssefi2018ImpactAorta, Stokes2023AneurysmalIndices}. However, a complete description of heart function and blood transport requires coupling throughout the entire cardiac cycle.

%This assumption is required because the flow information coming from Alya is, in principle, a defective condition: we have information on the flow rate but lack detailed pointwise velocity data.

%An alternative to our assumption is to adopt a Lagrange-multiplier approach in HemeLB, as described in \cite{veneziani2005flow}. This method presumes that the traction on the inlet section is normal to the surface and uniform across space, enabling the avoidance of any assumptions regarding the spatial distribution of the inlet velocity.

\rev{A fourth limitation is the assumption of a rigid aorta, which neglects its natural compliance. In reality, the aortic cross-section undergoes significant changes throughout the cardiac cycle, as demonstrated in both clinical \cite{metafratzi2002clinical, kuecherer2000evaluation, wang2023review} and computational studies, where fluid-structure interaction models are developed \cite{2009CardiovascularSystem, de2003three, sturla2013impact, bertagna2014model, flamini2016immersed}. This limitation affects the computed flow velocity and pressure, as arterial compliance plays a crucial role in dampening pressure waves and modulating blood flow dynamics. Within a lattice Boltzmann framework, fluid–structure interactions can be modelled by coupling the fluid dynamics with a finite element representation of the aortic wall mechanics, for instance, through the immersed boundary method \cite{Fringand2024AMethods, Wei2023DirectSimulations}.}

%% file: 5_Conclusion.tex
\section{Conclusion} \label{sec:conclusion}

We have successfully constructed a 3D model of the cardiovascular system by coupling the 3D electromechanical model of the heart in Alya with the 3D fluid mechanics model of vascular blood flow in HemeLB. Our file-based partitioned coupling scheme has been tested using idealised and realistic anatomies, demonstrating \rev{its} reliability and effectiveness. The implementation of a viscous sponge layer for blood flow and a low-pass filter at the coupling interface has proven effective in stabilising the coupled simulations. The coupling scheme is also efficient, requiring minimal additional computation time relative to advancing individual time steps in the heart and blood flow models.

\rev{Notably, our coupled model predicts muscle displacement and aortic wall shear stress differently than the standalone heart and blood flow models, highlighting the critical importance of coupling between cardiac and vascular dynamics to capture their intricate interactions in cardiovascular simulations. Muscle displacement shows peak differences of up to 22 \% at the endocardium, while TAWSS in the ascending aorta differs by as much as 20 \%. Furthermore, our simulations highlight the model's ability to capture key features of cardiac disease by assessing the impact of myocardial scarring on downstream vascular flow. This is made possible by the coupled interaction between the cardiac and vascular dynamics. As scar severity increases, both cardiac function and vascular haemodynamics deteriorate, with variations in TAWSS reaching 87 \% in the most severe case.}

Looking ahead, this coupled model can be further tested by applying it to a wide range of pathophysiological conditions, such as heart failure, hypertension, and congenital heart defects. Future work can also explore the integration of additional physiological systems, such as the respiratory and nervous systems, to enhance the fidelity of virtual human models and digital twins. Moreover, advancements in computational power and algorithms will likely allow for real-time simulations, thereby broadening the practical applications of these models in clinical decision making and personalised medicine.

This study exemplifies the successful multi-component coupling of models developed by separate research groups, underscoring the collaborative nature of scientific research. By bridging different modelling approaches, we have laid the groundwork for the development of virtual human models and digital twins which encompasses various biophysical processes within the human body. This integrative effort paves the way for more sophisticated and comprehensive simulations of human physiology, which will ultimately be essential in both research and clinical settings \cite{Coveney2023VirtualLife}.

%% file: 6_Appendix.tex
\section{Computational Performance of the Coupling Scheme} 
\label{sec:computational_performance}
We evaluate the computational performance of the coupling scheme by comparing the time spent on coupling with the time taken for one \rev{coupling step} in each solver during the coupling phase.

We record the elapsed time of each \rev{coupling iteration} in Alya and HemeLB and calculate their averages in the coupling phase. \rev{Here, one coupling iteration is equivalent to one time step in Alya and 16 time steps in HemeLB (since $k = 16$ in Test 2).} To determine the time spent on coupling, we measure the total time required to perform all procedures related to coupling, including the time for writing and reading data in files and the latency, and calculate its average. We perform these measurements using the problem outlined in Test 2.

\begin{figure}[!htb]
    \centering
    \begin{subfigure}{0.48\linewidth}
        \centering
        \includegraphics[width=\linewidth]{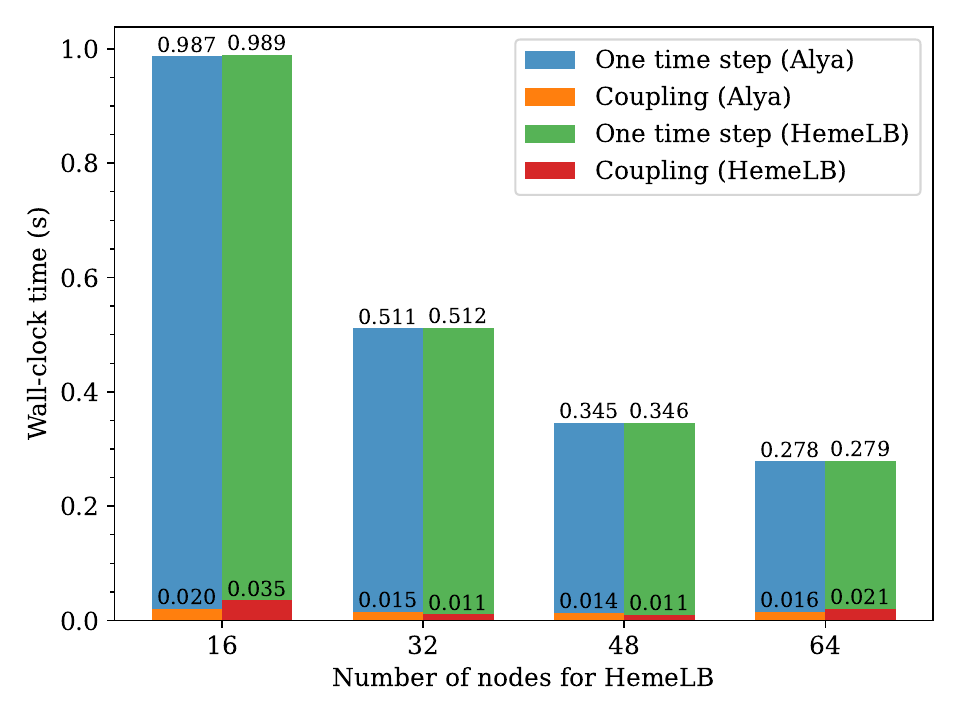}
        \caption{One compute node for Alya}
    \end{subfigure}
    \hfill
    \begin{subfigure}{0.48\linewidth}
        \centering
        \includegraphics[width=\linewidth]{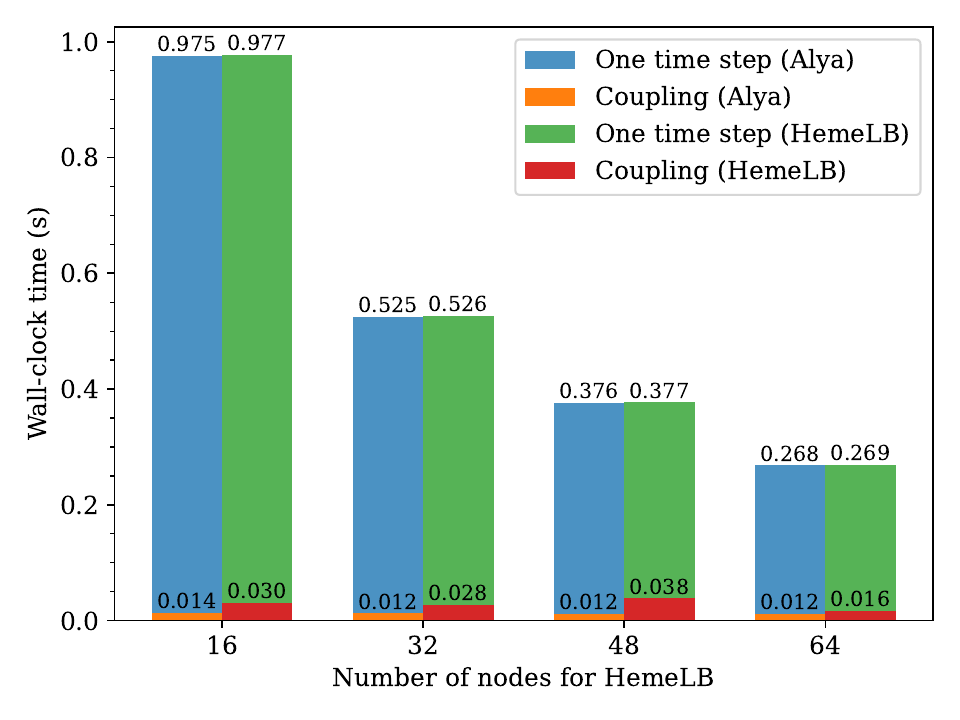}
        \caption{Two compute nodes for Alya}
    \end{subfigure}
    \caption{Average \rev{elapsed} time and coupling time in Alya and HemeLB \rev{per coupling iteration} using various numbers of compute nodes. The results are shown for Alya using (a) one compute node and (b) two compute nodes with each compute node consisting of 128 CPU cores.}
    \label{fig:measureTime}
\end{figure}

To understand the effects of parallelisation on the efficiency of the coupling scheme, we vary the number of CPU cores (or compute nodes) used by Alya and HemeLB. The results of the measurements are presented in Figure \ref{fig:measureTime}. While the average elapsed time for one \rev{coupling iteration} decreases with the number of compute nodes used for HemeLB, the average time spent on coupling in Alya and HemeLB remains relatively constant between 0.01 s and 0.04 s. For the largest number of nodes tested, specifically two nodes for Alya and 64 nodes for HemeLB, the percentage of the time spent on coupling relative to the elapsed time per \rev{coupling iteration} is about 4 \% for Alya and 6 \% for HemeLB.

\rev{In our current setup, the computational overhead from coupling is minimal because only a small amount of data is exchanged at each time step. Under these conditions, the file-based partitioned coupling approach remains efficient. However, in cases where large amounts of data need to be exchanged, such as full 3D fields in fluid–structure interaction or electromechanics simulations, using a dedicated coupling framework like MUSCLE3 \cite{Veen2020Easing3} or a monolithic software approach \cite{Santiago2018FullySupercomputers, fedele2023comprehensive, Augustin2021ACirculation, bucelli2023mathematical, zingaro2024electromechanics} may be more efficient. These approaches avoid repeated file I/O and allow for better exploitation of parallelism within the solver. Determining the point at which file-based coupling becomes inefficient depends on multiple factors, including data volume, exchange frequency, and solver architecture. Beyond performance, long-term maintainability and the planned longevity of the coupling framework also play a crucial role in selecting the most suitable approach. These computational and practical considerations together shape the choice of coupling strategy and represent an interesting direction for further investigation.}

%For Alya, we note that the difference between the elapsed times $\Tilde{t}_\text{elapsed}^\mathcal{A}$ in the first and second heartbeats represents the additional time due to the replacement of the WK2 model with HemeLB. By subtracting $\Tilde{t}_\text{elapsed}^\mathcal{H}$ from this difference, we obtain the time spent on the coupling in Alya. The average of this value is denoted by $\Tilde{t}_\text{coupling}^\mathcal{A}$.